\documentclass[fleqn,usenatbib]{mnras}
\usepackage{newtxtext,newtxmath}
\usepackage[caption=false]{subfig}
\usepackage[T1]{fontenc}
\usepackage{ae,aecompl}
\usepackage{graphicx}
\usepackage{amsmath}
\usepackage{siunitx}

\usepackage{xcolor}

\def\kms{\, \mathrm{km}\, {\mathrm s}^{-1}}

\title[The interacting binary V1315 Cas]{Everything that glitters is not gold: V1315 Cas is not a dormant black hole}

\author[Zak et al.]{
J. Zak$^{1,2}$\thanks{E-mail: jirizak1@seznam.cz},
D. Jones$^{2,3,4}$,
H. M. J. Boffin$^{1}$,
P.\,G. Beck$^{2,3,5}$,
J. Klencki$^{1}$,
J. Bodensteiner$^{1}$,
\newauthor 
T. Shenar$^{6}$,
H. Van Winckel$^{7}$,
M. Skarka$^{8,9}$,
K. Arellano-C\'ordova$^{10}$,
J. Viuho$^{11,4}$,
P. Sowicka$^{12}$,
\newauthor
E. W. Guenther$^{13}$, and
A. Hatzes$^{13}$
\\
$^{1}$ European Southern Observatory, 85748 Garching, Germany\\
$^{2}$Instituto de Astrof\'{\i}sica de Canarias, E-38200 La Laguna, Tenerife, Spain\\
$^{3}$Departamento de Astrof\'{\i}sica, Universidad de La Laguna, E-38206 La Laguna, Tenerife, Spain\\
$^{4}$ Nordic Optical Telescope, Rambla Jos\'e Ana Fern\'andez P\'erez 7, 38711, Bre\~na Baja, Spain\\
$^{5}$ Institut für Physik, Karl-Franzens Universität Graz, Universitätsplatz 5/II, NAWI Graz, 8010 Graz, Austria\\
$^{6}$ Anton Pannekoek Institute for
Astronomy, University of Amsterdam,
the Netherlands\\
$^{7}$ Institute of Astronomy, KU Leuven, Celestijnenlaan 200D, B-3001 Leuven, Belgium\\
$^{8}$ Department of Theoretical Physics and Astrophysics, Masaryk Univesity, Kotl\'a\v{r}sk\'a 2, 60200 Brno, Czech Republic\\
$^{9}$ Astronomical Institute of the Czech Academy of Sciences, Fri\v{c}ova 298, CZ-25165 Ond\v{r}ejov, Czech Republic\\
$^{10}$ Institute for Astronomy, University of Edinburgh, Royal Observatory, Edinburgh, EH9 3HJ, UK\\
$^{11}$ DAWN, Niels Bohr Institute, University of Copenhagen, Jagtvej 128, 2200 Copenhagen N, Denmark\\
$^{12}$ Nicolaus Copernicus Astronomical Center, Bartycka 18, PL-00-716 Warsaw, Poland\\
$^{13}$ Thueringer Landessternwarte Tautenburg, Sternwarte 5, 07778 Tautenburg, Germany
}
\date{Accepted XXX. Received YYY; in original form ZZZ}
\pubyear{2020}
\begin{document}
\label{firstpage}
\pagerange{\pageref{firstpage}--\pageref{lastpage}}
\maketitle

\begin{abstract}
The quest for quiet or dormant black holes has been ongoing since several decades. Ellipsoidal variables possibly indicate the existence of a very high-mass invisible companion and are thought to be one of the best ways to find such dormant black holes. This, however, is not a panacea as we show here with one example. 
We indeed report the discovery of a new semi-detached interacting binary, V1315 Cas, discovered as an ellipsoidal variable. Using data from photometric surveys (ASAS-SN, TESS) and  high-resolution spectroscopy, we derived a nearly circular orbit with an orbital period of $P_{\rm{orb}}$=34.54 d. The binary system consists of an evolved F-type star primary that is likely still filling its Roche lobe and a B-type star secondary. Using \textsc{phoebe}2, we derived the following masses and radii: for the primary, $M_p =0.84 \pm 0.03 \,\rm{M}_{\sun}$ and $R_p =18.51^{+0.12}_{-0.07} \, \rm{R}_{\sun}$; for the secondary, $M_s =7.3 \pm 0.3 \,\rm{M}_{\sun}$\, and $R_s =4.02^{+2.3}_{-2.0}\,\rm{R}_{\sun}$. Modeling the evolution  of the system with MESA, we found an age of $\sim$\SI{7.7e7} years. The system is at the end of a period of rapid non-conservative mass transfer that reversed its mass ratio, while significantly widening its orbit. The primary shows carbon depletion and nitrogen overabundance, indicative of CNO processed material being exposed due to mass transfer.  An infrared excess as well as stationary H$\alpha$ emission suggest the presence of a circumstellar or circumbinary disc. V1315 Cas will likely become a detached stripped star binary.

\end{abstract}


\section{Dormant black holes}

The demographics of black holes (BHs) is severely limited as we currently know (to varying degrees of confidence) only around 100 BH systems. The vast majority of these are found in binary systems and were discovered 
in x-ray surveys. Only a handful of objects are known to be strong candidates to contain x-ray quiet BHs \citep[or ``dormant'';][]{khok18,gie18,shenBH22,mah22,EB23BH}. Assuming that only a fraction of BHs are actively accreting \citep{kalo04}, dormant BHs should
be populous in our Galaxy -- for example, \citet{mcc06} estimated the population of BHs in our Galaxy to be around \SI{3e8}\  -- and their absence has been discussed for
several decades \citep{tri69}.  
Recently, several studies claimed the discovery of a quiescent stellar-mass black hole. 
However, subsequent studies \citep{irr20,she20,bod20,elb_q21,elbd22_ngc} have shown that such systems do not actually contain a stellar-mass black hole, but are the outcome of binary evolution that produced a stripped star with a Be star companion. A summary of recent developments can be found in \citet{bod22}.

A proper description of the black hole population and
stellar evolution understanding requires a dramatic increase of the number of
known BHs. \citet{gom21} suggested that ellipsoidal variables with large photometric amplitude could be used to detect dormant BHs.
This triggered our interest to study in detail the system V1315 Cas.

V1315 Cas 
is a bright object ($V$=10.3), originally classified as a Cepheid variable star in the Variable Stars Index catalogue \citep[VSX,][]{wat06}. We observed this target during our study of low-amplitude Cepheids but we quickly realized that this star was misclassified. Our preliminary analysis indicated large ellipsoidal variations and a low contribution to the optical flux from the secondary, making it a perfect candidate for hosting a compact object. This was our hypothesis for a long time and the reason why we embarked on a detailed study of this object. Only a combination of thorough spectroscopic and photometric analysis as presented in this paper revealed the true nature of the system -- a stripped star binary system.

\section{Data sets}
\subsection{Photometric data}

V1315 Cas was observed as part of the ASAS-SN survey \citep{koch17,jay18}, providing $V$- and $g$-band photometry. Observations span four observing seasons in the $V$ filter and four additional seasons in the $g$ filter. The unfolded light curve is shown in Fig. \ref{fig:lcun} in the Appendix, while the phase-folded $V$-band light curve is presented in Fig. \ref{fig:lcv}.

\begin{figure}
 \includegraphics[width=\columnwidth]{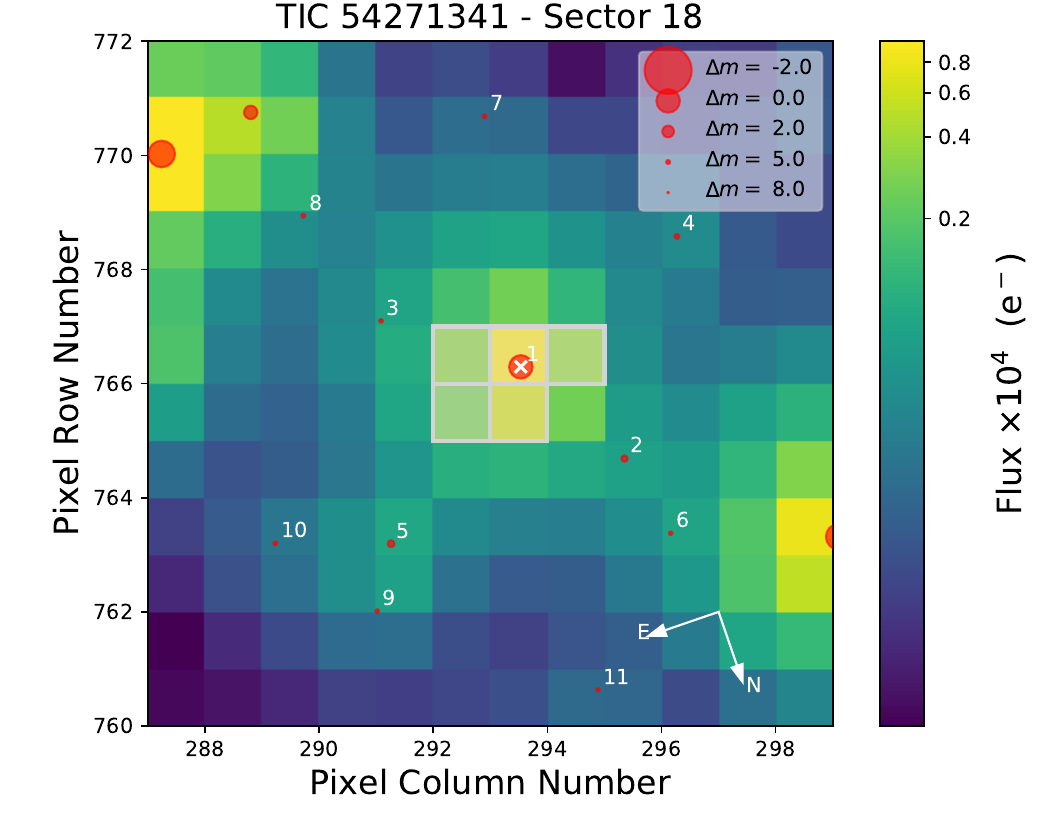}
 \caption{TESS Target Pixel File plot of V1315 Cas (TIC54271341) created using \texttt{tpfplotter}.}
 \label{fig:tpf}
\end{figure}

\begin{figure}
 \includegraphics[width=\columnwidth]{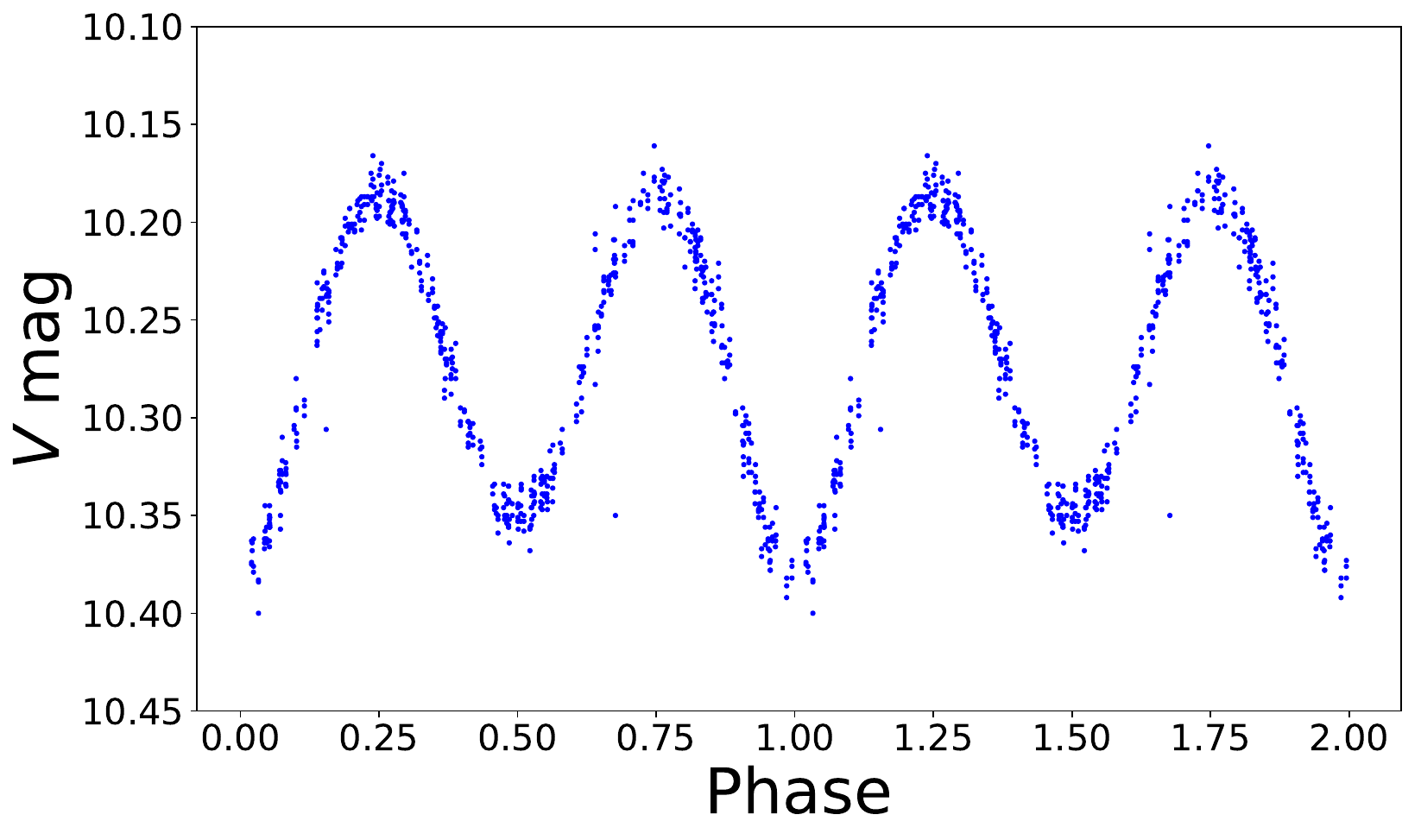}
 \caption{\textit{V}-band light curve with data from ASAS-SN, phased with the orbital period, $\rm{P}$=34.5352 days. Data points are repeated for the second cycle for better visualisation.}
 \label{fig:lcv}
\end{figure}

These data were complemented by photometric data from NASA's \textit{Transiting Exoplanet Survey Satellite}  \citep[TESS;][]{rick15} in the wavelength range 600 to 1000 nm. V1315 Cas was observed in sector 18 with a 30 minute cadence -- the Target Pixel File shown in Figure \ref{fig:tpf} proves that there are only faint nearby stars that do not impact the light curve. In addition, these stars are not in the aperture which is highlighted by the white squares. The differential magnitude of the source, with respect to several nearby and isolated comparison stars of similar brightness, was extracted from the full frame images (FFIs) via simple aperture photometry with an aperture radius of 2 pixels. The resulting phase-folded light curve is shown in Fig.  \ref{fig:phob}.

\subsection{Spectroscopic data}
\label{sdata}
Spectroscopic observations were first carried out using the echelle spectrograph ($R=~35\,000$) at the 2-m Alfred Jensch Telescope in Tautenburg, Germany, covering a spectral range from 465 to 759 nm \citep{hat03}. 
A radial-velocity (RV) monitoring campaign was then started using the HERMES \citep{Raskin2011} 377-900~nm echelle spectrograph ($R=~85\,000$) at the 1.2-m Mercator telescope located at La Palma, Spain. 

The Tautenburg data were reduced using standard procedures in IRAF. The HERMES data were reduced using the dedicated automated data reduction pipeline (HermesDRS) version 7.0. The signal-to-noise ratios (S/N) of the acquired spectra range from 20 to 45 for the HERMES spectra and are about 50 for the Tautenburg spectra. The dates of the individual observations can be found in Table \ref{tab:rv}. The exposure times varied between 600 s and 1200 s. 

Additionally, a low-resolution ($R=~350$) flux calibrated spectrum was obtained with the ALFOSC instrument mounted at the 2.5~m Nordic Optical Telescope \citep[NOT;][]{2010ASSP...14..211D}. The spectrum was taken on September 5, 2020, with a 1\arcsec\ wide slit and the grism \#3 centered at 432~nm, allowing a wavelength coverage from 330 to 705~nm. The exposure time was 240~s. The spectrum was reduced using standard \textsc{starlink} routines \citep{star}.

\section{Analysis}

\subsection{Astrometry}
\label{gaiainfo}

V1315 Cas is listed in the Gaia DR3 catalogue (ID:523184053818900224) with a parallax $\varpi=0.6967 \pm 0.0148$ mas, which, when correcting for the zero point discussed in \citet{Lin21} leads to a distance of $d=1.382^{+0.029}_{-0.028}$ kpc.

In addition, the Gaia EDR3 \citep{lin20,gai20} proper motions and parallaxes suggests that V1315 Cas is not part of any stellar cluster.

The Renormalised Unit Weight Error (RUWE\footnote{https://www.cosmos.esa.int/web/gaia/dr2-known-issues}) coefficient $\varrho$ returned in Gaia EDR3 is $\varrho = 1.219$, smaller than the canonical value of 1.4, which is usually meant to suggest binarity and/or unreliable astrometric solutions. This is likely no surprise given the very short orbital period.

\subsection{Light curves}
\label{sec:lc}

The photometric data of V1315 Cas show a roughly sinusoidal variability with an amplitude of approximately 0.2 mag in all bands (Fig.~\ref{fig:phob}). 

The depths of successive minima are different by approximately 0.03\,mag, which, in combination with constant maximum brightness, strongly indicate ellipsoidal variability and two minima per orbital period. Such variations can also be obtained by spots \citep[see e.g. fig. 7 in][]{ska22}. However, the binary nature of the variations is obvious from our findings presented in sect. \ref{sec:rvs}. Previous analyses did not identify the different minima \citep[e.g.,][]{sokolovsky14} and, as such, concluded that the period was roughly half the true period.  The HJD ephemeris of the deepest minimum as determined from the ASAS-SN $V$ curve, which has the longest temporal coverage, $\approx$1400 days, is 
\begin{equation}
2457037.0204 (0.0184) + 34.5352 (0.0006) E.
\end{equation}

The data from ASAS-SN appear to show systematically underestimated uncertainties of the observed magnitudes compared to the scatter in the data. This may be due to the fact that the ASAS-SN data are in the saturation regime below $V < 11$ mag, even if they do not show any systematics typical of saturation (priv. comm.).

 \subsection{Spectral analysis}
 \label{sec:rvs}

  \subsubsection{Stellar parameters}
\label{sec:spec}

We determined the stellar parameters of the primary star\footnote{In this paper, we refer to the F-type evolved star as the primary as it is the more luminous in the visible region, even if we will show later that it is the less massive.} using \textsc{iSpec} \citep{blan14,bla19}, specifically the spectral synthesis method with the MARCS atmosphere models and the SPECTRUM code. We considered the spectral region from 660 to 760 nm where the light contribution from the hot companion is minimal. We selected only Fe I and Fe II lines for the analysis. The obtained stellar parameters are presented in Table \ref{params}. We compared the determined temperature to the temperature provided by \textit{Gaia} DR2 of $T_{\rm{eff}}$=4376\,K which is significantly lower than our value of $T_{\rm{eff}}$=7000\,K. A possible explanation for this disagreement could be dust, large interstellar extinction and the binary nature of the system. We show in Fig \ref{speccomp} a comparison of our best-fit derived spectrum to one using the temperature derived by \textit{Gaia} DR2, clearly demonstrating that the Gaia temperature is not reliable. To assess how the light contamination from the unaccounted light of the secondary affects the spectrosocpic parameters, we generated synthetic spectrum with the above derived parameters of the primary, matched the S/N of the observed spectra, polluted the spectrum and obtained the spectroscopic parameters again. We found additional uncertainities might affect the derived parameters up to 200 K, 0.2 in surface gravity and 0.1 for metallicity. Furthermore, we detect several prominent diffuse interstellar bands (DIBs) at 578.02 nm, 579.68 nm, and 661.32 nm. By measuring their equivalent widths and using the relation of \citet{herb95}, we obtain $A_V$ $\approx$ 2.7. This is slightly higher than the value $A_V$ $\approx$2 suggested by the Pan-STARRS 1 dust map \citep{gre19}, but not surprising given the large distance of V1315 Cas. By contrast, \textit{Gaia} DR2 quotes a much lower value of $A_g=0.49$, corresponding to $A_V=0.57$. In the analysis, the strong interstellar sodium lines were not used to infer the extinction.

\begin{figure}
\captionsetup[subfloat]{farskip=0pt,captionskip=1pt}
{%
  \includegraphics[clip,width=\columnwidth]{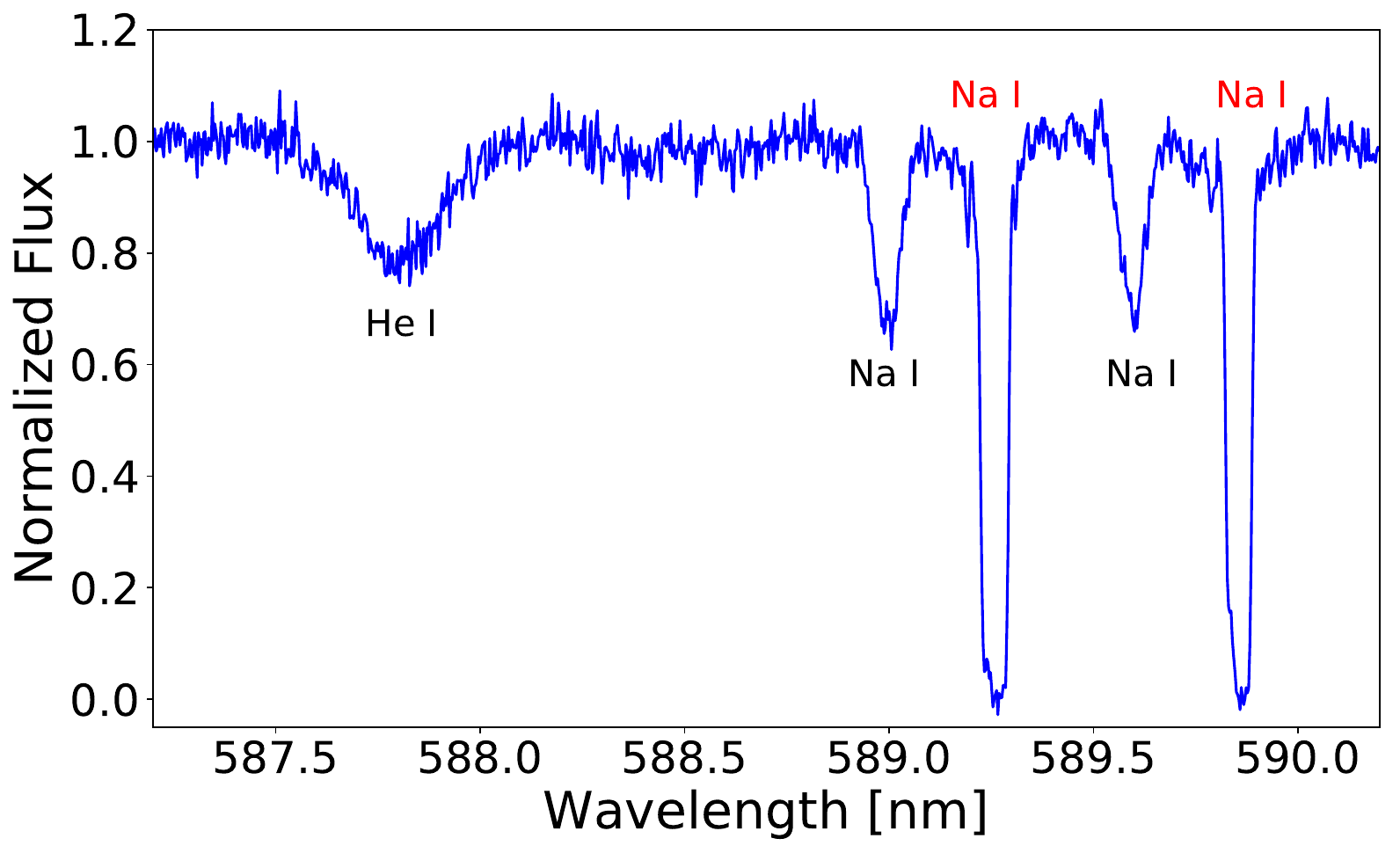}
}

{%
  \includegraphics[clip,width=\columnwidth]{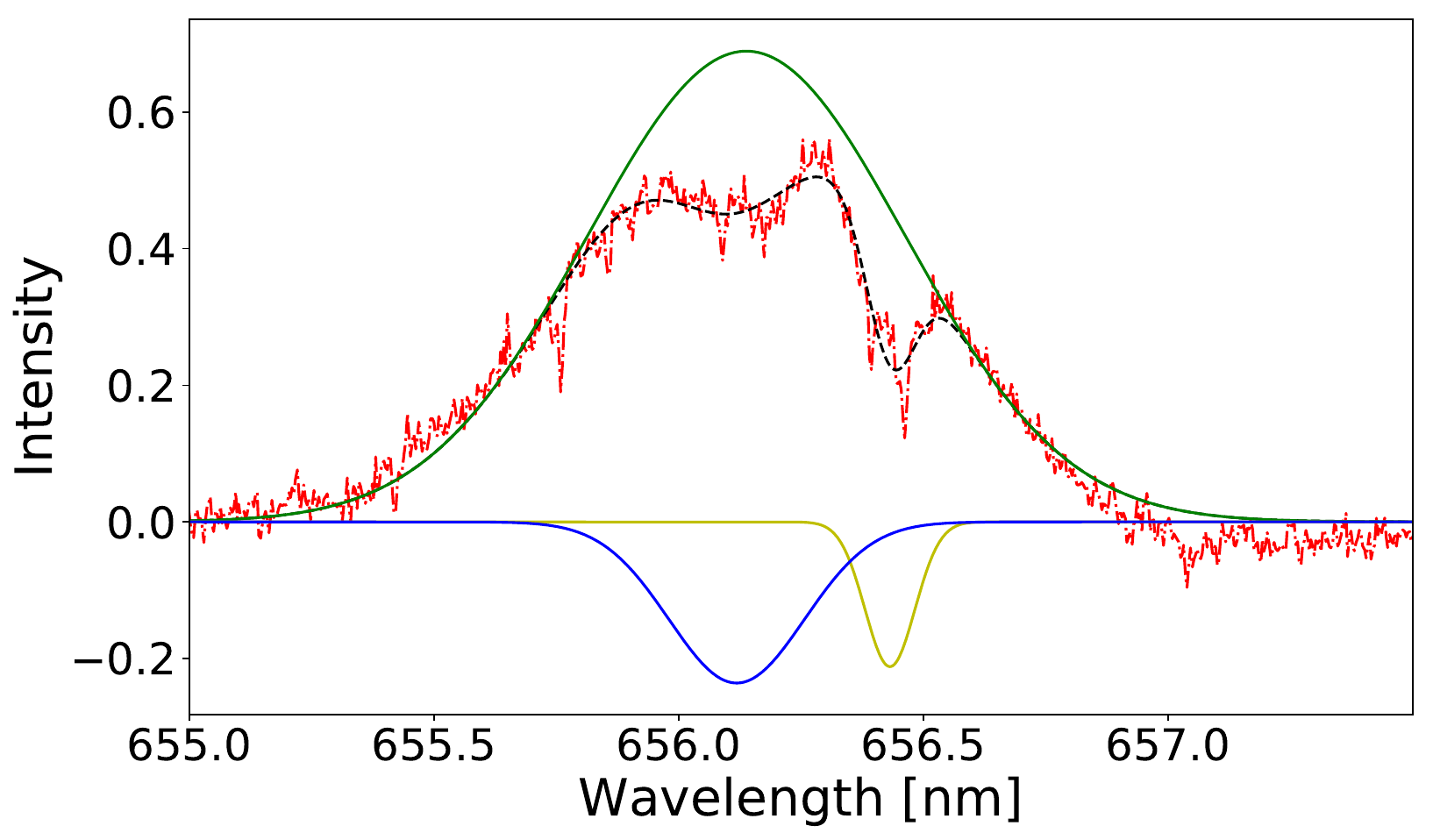}%
}

\caption{\textit{Top:} Part of the HERMES spectrum around the 587.5 nm neutral helium line and pairs of stellar (black) and interstellar (red) sodium lines. The different $v \sin i$ implied by these lines reveal the existence of two stars. \textit{Bottom}: The complex structure of the H$\alpha$ line. The observed spectrum is shown in red. The best fit using 3 components is overplotted in black. The dominant emission component is shown in green, while the absorption components are displayed in blue and yellow.}
\label{fig:haspec}
\end{figure}

\begin{table}
\caption{Stellar parameters of the primary determined from spectroscopic analysis.}
\centering
\begin{tabular}{ c  c  c  c c }
\hline
$T_{\rm{eff}}$ (K)& $\log g$ & [Fe/H] &$\xi_t$ (km/s)  &$v\sin{i}$ (km/s)  \\
\hline  
  7000 $\pm$ 130 &1.8$\pm$ 0.15 & 0.1$\pm$ 0.1 & 1.05 $\pm$ 0.4 & 25.8 $\pm$ 1.9 \\
\hline 

\end{tabular}
\label{params}
\end{table}
 
 \subsubsection{Radial velocities}
The observed spectra show a number of features, including clear Doppler-shifted narrow metal lines as a function of the orbital phase. Moreover, the H$\alpha$ line presents a complex structure. Using Gaussian fitting, we were able to distinguish three distinct components: two absorption lines and one static emission component (Fig.~\ref{fig:haspec} and \ref{fig:hal}). Furthermore, broad neutral helium lines at 587.56 and 667.81 nm are present alongside the narrow metal lines (Fig. \ref{fig:haspec}). As neutral helium absorption lines become visible at higher temperatures compared to the observed metal lines, they are unlikely to originate from the same source, confirming the binary nature of the system and the absence of a BH. 

\begin{figure}
 \includegraphics[width=\columnwidth]{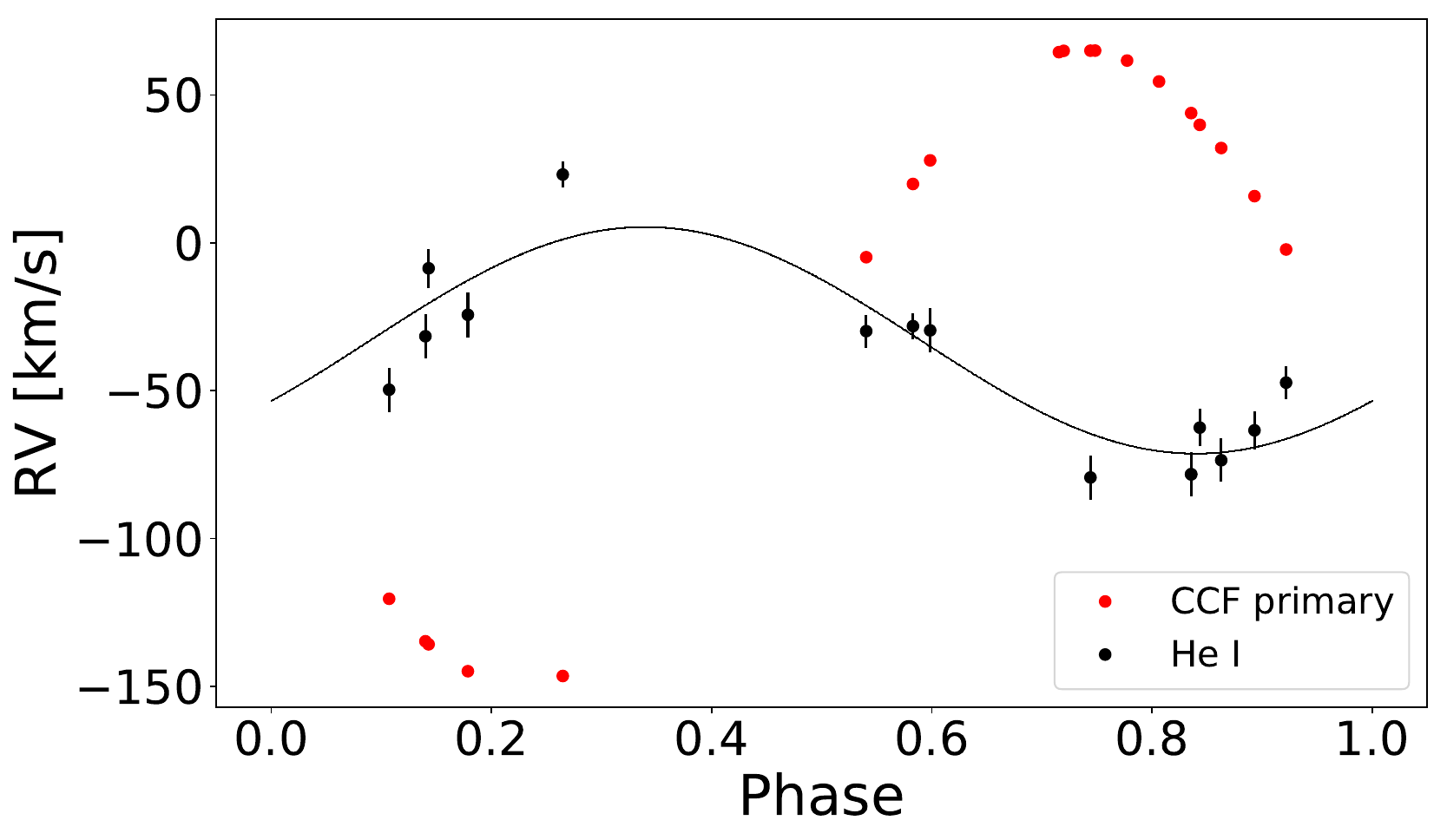}
 \caption{Radial velocities of two components in the system. The cross correlation function (CCF) was used to obtain the RVs of the primary (red). Gaussian fitting of the 587.5 nm helium I line was used to obtain the RVs of the secondary (black). Fit to the radial velocities of the secondary is shown with the black line. See text for more details.}
 \label{fig:rv2comp}
\end{figure}

The radial velocities of the primary star were determined via cross-correlation with a template spectrum with parameters from Table \ref{params} in \textsc{iSpec}, excluding regions with Balmer, telluric and other spectral lines that likely do not originate from the primary star. 
In addition, we used a Gaussian fit of the 587.5 He I line to derive the RVs. The obtained radial velocity curves, folded on the orbital period derived in section \ref{sec:lc}, are shown in Fig. \ref{fig:rv2comp}. One absorption component of the H$\alpha$ line has the same RVs as the primary star. The second component of the H$\alpha$ line has RVs similar to the helium lines. The individual radial velocity measurements of the object are listed in Table \ref{tab:rv}.

\begin{figure}
 \includegraphics[width=\columnwidth]{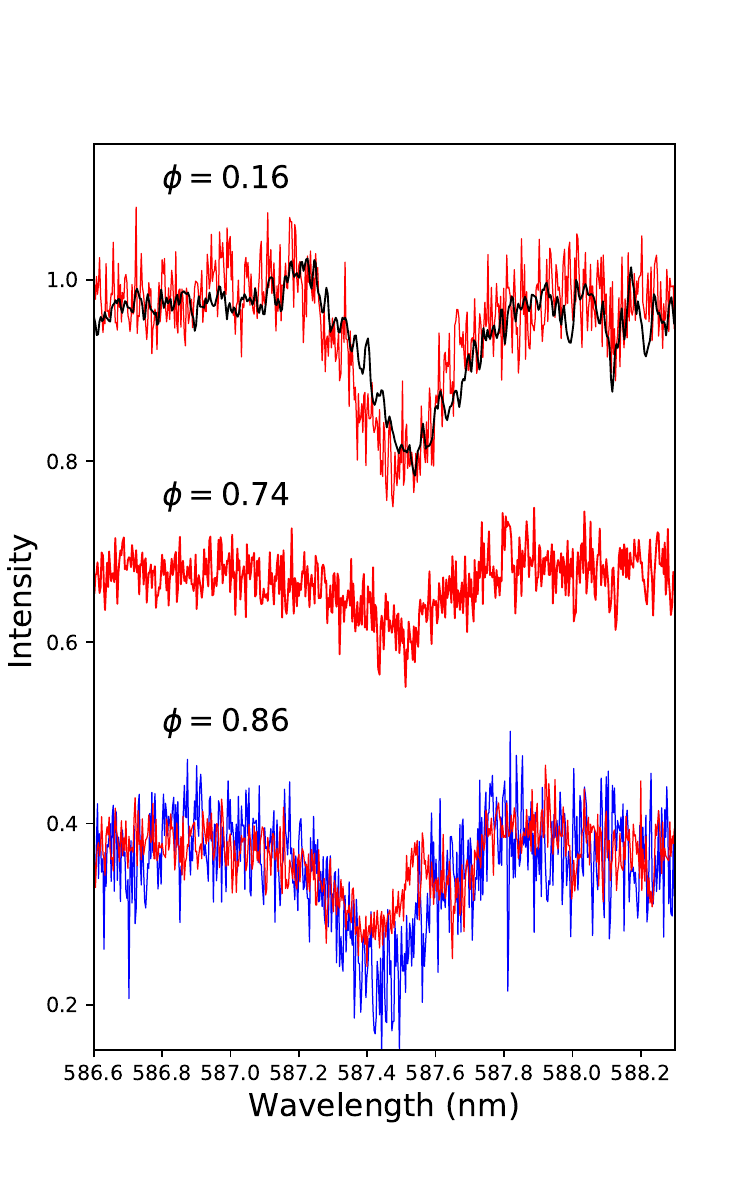}
\caption{Line profile variations of the $\lambda$587.5\,nm He I line at different orbital phases. In the two cases where we obtained spectra at the same orbital phase in different observing seasons, the distinct shapes of the features suggest a variability origin different from the orbital motion. For visualization purposes, the spectra at different orbital phases are offset vertically by 0.3 units.}
 \label{fig:tycheli}
\end{figure}

\subsubsection{Helium lines}
\label{hellines}
As mentioned above, the observed spectra reveal the presence of neutral helium lines at 587, 667 and 706 nm. Furthermore, we see hints of helium I lines at bluer wavelength at 501.5 and 447.1 nm, but those lines are severely blended and weak hence not useful for further study. We attempted to derive stellar parameters from fitting or comparing synthetic spectra to the observed helium lines. However, the observed spectra do not have sufficient signal-to-noise ratio and are hindered by variability described below to produce meaningful results from the helium lines. We can only rule out synchronous rotation of the secondary (rotation velocity corresponding to $\sim$ 5km/s) as our analysis yields projected rotation velocity between 60 and 300km/s in various epochs. Such discrepancy is likely the results of the variability of the helium lines. Further analysis was performed on the 587 nm helium I line as the contamination from the primary is presumably negligible.
Our spectra cover three observing seasons and we observe long ($>> P_{\rm{orb}}$) and short ($\sim P_{\rm{orb}}$) variability in the observed helium I lines (Fig. \ref{fig:tycheli}). This includes equivalent width changes and appearance of line asymmetries. These asymmetries can be of various origins. \citet{barr13} attributed similar asymmetries to a filling emission from a higher local temperature regions such as hot or bright spots.

\subsection{Chemical abundances}
\label{abund}

We have determined the chemical abundances of V1315 Cas via spectral synthesis in \textsc{iSpec}, using  MARCS model atmospheres \citep{gus08}. To increase the S/N, we have stacked three spectra that were obtained near the same orbital phase to minimize line profile distortions. We present the obtained results in Table\,\ref{abu}. We report strong enhancement of nitrogen in the primary star alongside with carbon depletion most likely due to the CNO cycle. This process converts carbon to nitrogen in the interior of intermediate massive stars. Given the large variations of the line profiles of the secondary over various timescales (sec. \ref{hellines}) we do not try to derive abundances of the secondary.

\begin{table}
\caption{Determined chemical abundances of the primary. }
\centering
\begin{tabular}{ c  c c c  c c c c c c c }
\hline
C & [N/H] & [O/H]   \\
\hline  
depleted & 1.51 $\pm$ 0.1 & -0.7 $\pm$ 0.15 \\
\hline 

\end{tabular}
\label{abu}
\end{table}

\subsection{Stellar parameters from simultaneous light and radial velocity curve modelling}
\label{phmod}

The parameters of the binary components were estimated via simultaneous modelling of the TESS, ASAS-SN \textit{V} and \textit{g} light and RV curves using the \textsc{phoebe}2 code \citep{prsa16,horvat18,jones20}.  Based on the spectroscopic analysis presented in section \ref{sec:spec}, the temperature of the primary star was fixed to 7000~K, while its mass and radius were allowed to vary. The binary inclination and the mass of the secondary were allowed to vary freely.

\begin{figure*}
 \includegraphics[width=\textwidth]{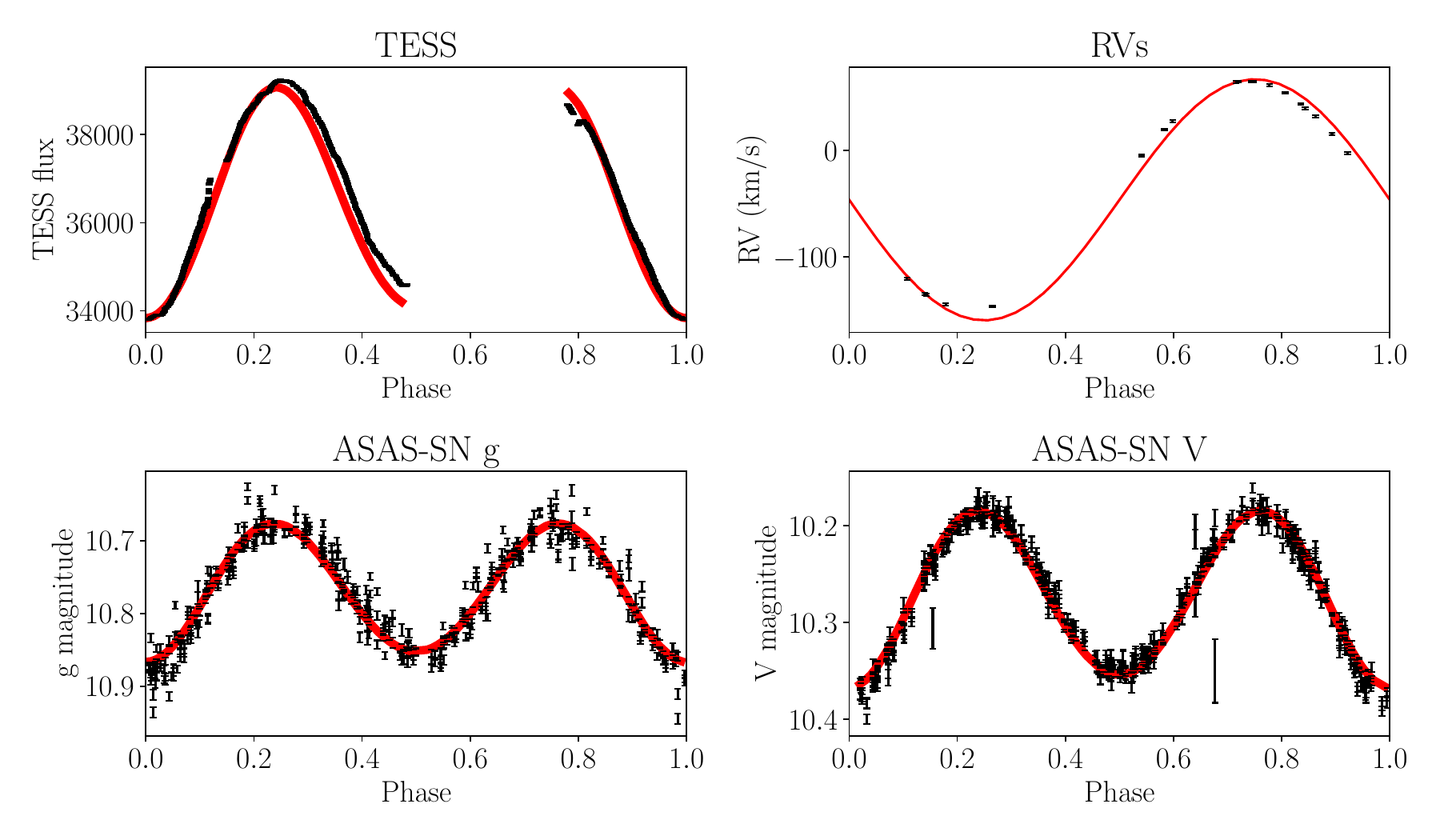}
 \caption{\textsc{phoebe} model simultaneously fitting the TESS and ASAS-SN light curves along with the observed radial velocities.  The best fitting model, for the observed light and radial velocity curves, is shown underlaid red.}
 \label{fig:phob}
\end{figure*}

In our model set-up we did not use any constrains on the system architecture (the system was modelled as a detached one). Atmospheres of both stars were modelled with Castelli \& Kurucz atmospheres and \textsc{phoebe2's} interpolated limb-darkening. The primary had a fixed albedo of 0.6, and the secondary 1.0. Furthermore, the gravity darkening of the primary
was fixed to be 0.32 and 1.0 for the secondary. The parameter space was explored using a Markov chain Monte Carlo (MCMC) method as described in \citet{boffin18} and \citet{jones19}. The parameters of the system determined by this analysis are shown in Table \ref{tab:phoebe}, while the interdependencies of the various parameters are shown in the form of a corner plot of the MCMC posteriors in Figure \ref{fig:mcmc}.  The best-fitting model light and radial velocity curves are shown overlaid on the data in Figure \ref{fig:phob}.
In our model we also fitted for a dilution ("3rd light") that was allowed to vary between 0 and 20\% of the total light. The final fit has 9\% $\pm$ 1\% third light.

\begin{table}
\caption{Binary and stellar parameters determined from simultaneous light and radial velocity curve modelling in PHOEBE.  See Figure \ref{fig:mcmc} for a corner plot of the fit posteriors.}
\centering
\renewcommand{\arraystretch}{1.5}
\begin{tabular}{ r r l}
\hline
Parameter & \multicolumn{2}{c}{Value}\\
\hline
Temperature of the primary (K) & 7000 & fixed\\
Mass of the primary (M$_\odot$) & 0.84 & $\pm$0.03\\
Radius of the primary (R$_\odot$) & 18.51 & $^{+0.12}_{-0.07}$\\
Surface gravity of the primary, log g & 1.83 & $^{+0.01}_{-0.01}$\\
Mass of the secondary (M$_\odot$) & 7.3 & $\pm$0.3\\
Radius of the secondary (R$_\odot$) & 4.02 & $^{+2.3}_{-2.0}$\\
Surface gravity of the secondary (log g) & 4.09 & $^{+2.3}_{-2.0}$\\
Temperature of the secondary (K) & 17290 & $^{+60}_{-40}$\\
Binary inclination (degrees) & 75.54& $^{+1.73}_{-1.28}$\\
RV-semi amplitude of the primary (km~s$^{-1}$) & 108.4 & $\pm$0.9\\
Mass function (M$_\odot$) & 4.6 & $\pm$0.1\\
Systemic velocity, $\gamma$ (km~s$^{-1}$) & $-46.26$& $^{+0.60}_{-0.35}$\\
Mass ratio & $0.1148$& $^{+0.0013}_{-0.0016}$\\[1mm]
\hline 
\end{tabular}
\label{tab:phoebe}
\end{table}

 \begin{figure}
 \includegraphics[width=\columnwidth]{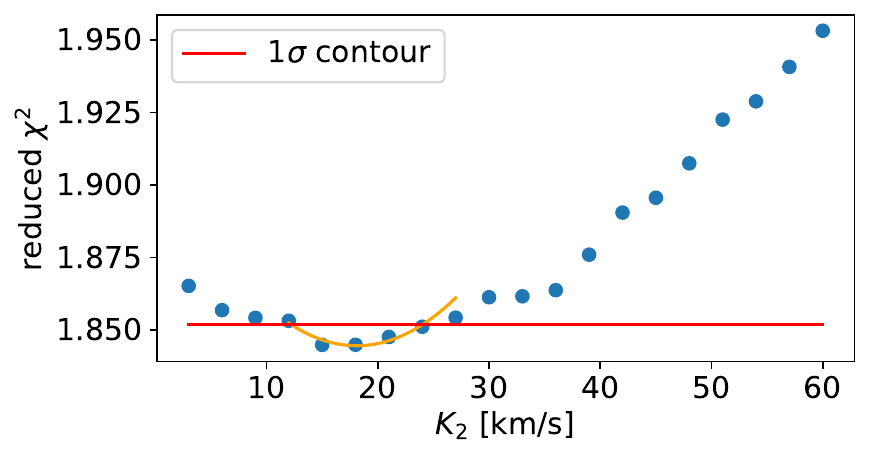}
 \caption{$\chi^2(K_2)$ map from disentangling of the H$\beta$ line. The formal minimum and error obtained are $K_2 = 18\pm 6\,\kms$.  }
 \label{fig:Hbetachi2}
\end{figure}

 \begin{figure}
 \includegraphics[width=\columnwidth]{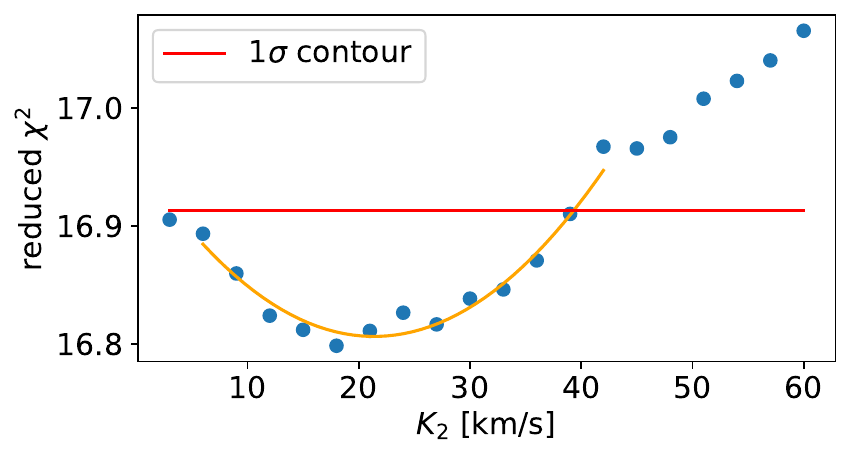}
 \caption{$\chi^2(K_2)$ map from disentangling of four He\,{\sc i} lines simultaneously: He\,{\sc i}\,$\lambda 433.8, 447.1, 587.6, 667.8$ nm. The formal minimum and error obtained are $K_2 = 21\pm 18\,\kms$.  }
 \label{fig:Hechi2}
\end{figure}

 \begin{figure}
 \includegraphics[width=\columnwidth]{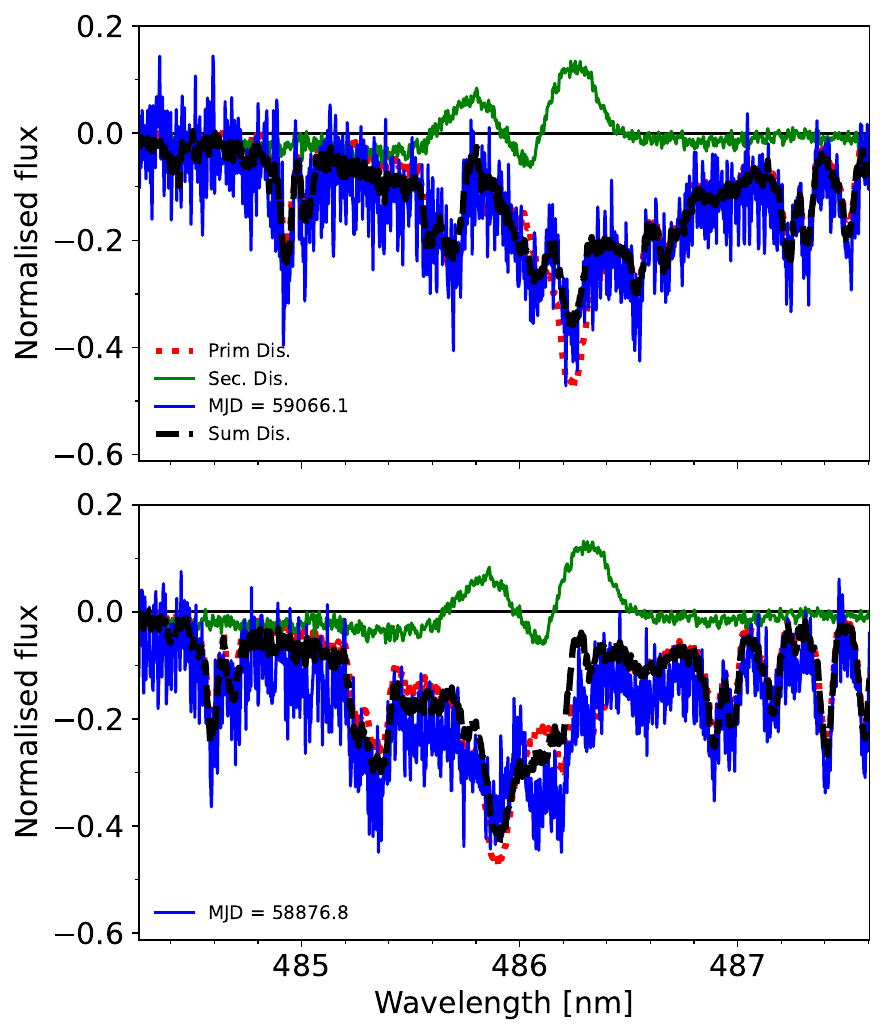}
 \caption{Disentangled spectra of the H$\beta$ region at the two RV extremes, as obtained for $K_2=18\,\kms$.  Shown are the  individual component spectra (red and green; Doppler-shifted according to the epochs shown), their sum (black), and the observations (blue). The MJDs are given in legend. }
 \label{fig:HbetaDis}
\end{figure}

 \begin{figure}
 \includegraphics[width=\columnwidth]{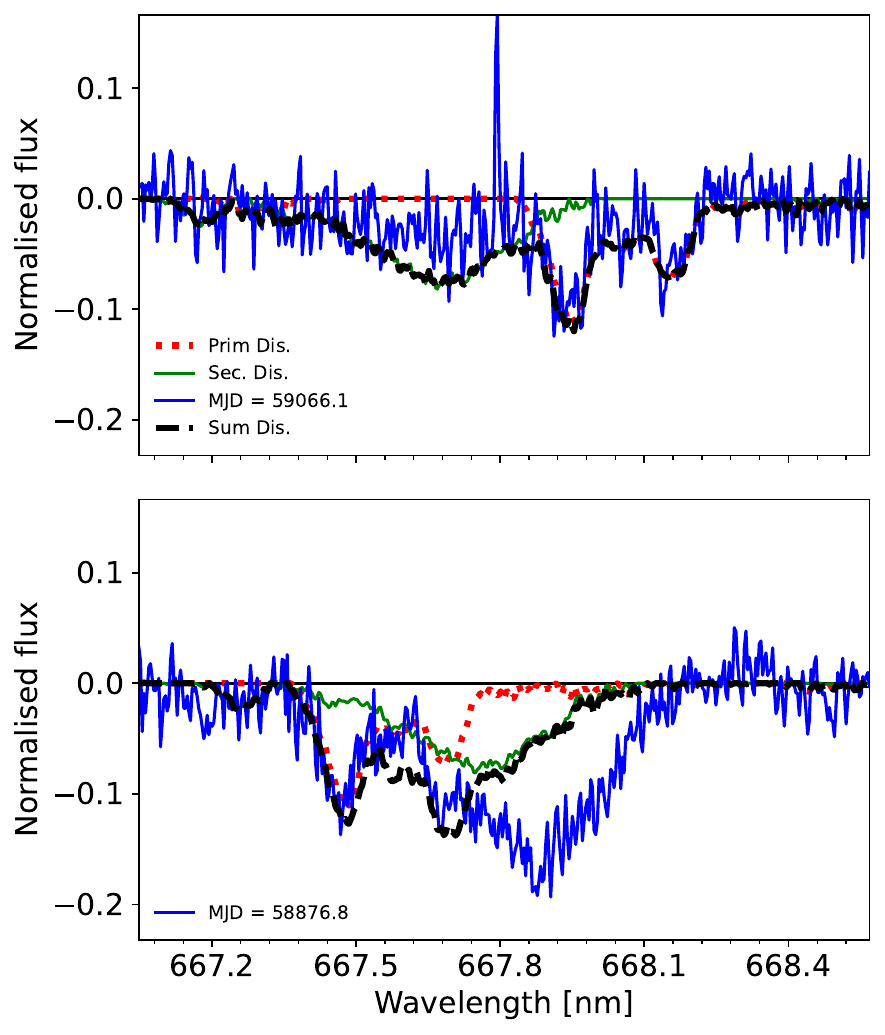}
 \caption{As Fig.\,\ref{fig:HbetaDis}, but for He\,{\sc i}\,$\lambda 667.8$ nm. The observation in the bottom panel shows a strong red-shifted absorption excess compared to the upper panel, implying a strong variability potentially stemming from circumstellar or circumbinary material. Since the disentangling procedure assumes constant line profiles, this feature cannot be reproduced. }
 \label{fig:HeDis}
\end{figure}

\subsection{Spectral disentangling}
\label{specdis}

We disentangle the observed composite spectra using the shift-and-add technique \citep[][]{Marchenko1998, Gonzalez2006, Shenar2017}. In short, the shift-and-add technique is an iterative procedure that uses all the observed spectra to compute the component spectra that best reproduce the observations. For a given set of orbital parameters, the RVs of the components at each epoch are known. Provided, say, the disentangled spectrum of the primary in the $i^{\rm th}$ iteration, the disentangled spectrum of the secondary in this iteration is computed by subtracting the primary spectrum from all observations (using the known RVs) and co-adding the residual observations in the frame-of-reference of the secondary. In the first iteration, a flat spectrum is assumed for the secondary star. The procedure typically converges within 50-100 iterations. The light ratio of the components needs to be adopted, and only impacts the final scaling of the disentangled spectra. As all orbital parameters except of $K_2$ are known, we fix the orbital parameters, but let $K_2$ vary, performing a grid disentangling. We hereby focus on different wavelength regions, namely the Balmer lines and the main \ion{He}{i} lines, and compute the $\chi^2$ of the match between observations and disentangled spectra for each $K_2$. For more information, we refer the reader to \citet{she20}.

Figure\,\ref{fig:Hbetachi2} shows the $\chi^2(K_2)$ map obtained for the  H$\beta$ region, implying $K_2 = 18\pm6\,\kms$. Figure\,\ref{fig:Hechi2} shows a similar plot for the four He\,{\sc i} lines at $\lambda 433.8, 447.1, 587.6, 667.8$, with a formal solution of $K_2 = 21 \pm 18\,\kms$. A weighted mean yields $K_2 = 18\pm6\,\kms$ (dominated by H$\beta$). Figures \ref{fig:HbetaDis} and \ref{fig:HeDis} show a comparison between the observations at RV extremes and the disentangled spectra obtained for the H$\beta$ and He\,{\sc i}\,$\lambda 667.8$ lines, respectively, when using $K_2 = 18\,\kms$. A light contribution of 30\% was adopted for the secondary. The H$\beta$ region is well reproduced, though discrepancies are apparent close to the core of the H$\beta$ line. Disentangling implies that the spectrum of the secondary is dominated by a double-peaked emission, reminiscent of a disk. 

The match between observations and disentangled spectra for the He\,{\sc i}\,$\lambda 667.8$ line (Fig.\,\ref{fig:HeDis}) is relatively poor and the reason is clear. The line shows very strong line-profile and equivalent-width variability, with a strong red-shifted absorption seen as the narrow-lined primary is receding from the observer. A similar behaviour is apparent in the other \ion{He}{i} lines such as $\lambda 587.5$.

Disentangling of the H$\alpha$ line results in a good match (comparable to Fig.\,\ref{fig:haspec}), but interestingly, the $\chi^2$ analysis implies $K_2$ values smaller than $\approx 10\,\kms$. This may suggest that the H$\alpha$ emission is dominated by a stationary disk (e.g.,  a circumbinary disk). However, it is possible that non-Doppler variability impacts these results, and such a conclusion should be taken with caution. 

In general, the disentangled spectrum of the secondary is dominated by broad Balmer and He\,{\sc i} absorption lines, which clearly originate in a stellar source.

\subsection{Spectral Energy Distribution}

We performed a Spectral Energy Distribution (SED) analysis using two data sets. The first one is the low-resolution flux-calibrated spectrum described in section \ref{sdata}. The second data set is broadband photometry covering a region from 0.43 to 22 \textmu m. We used the $B_T$, $V_T$ magnitudes from \textit{Tycho-2}, \textit{BVgri} magnitudes from APASS, the \textit{JHKs} magnitudes from 2MASS, the \textit{W}1-\textit{W}4 magnitudes from WISE, and the \textit{G} magnitude from Gaia.

\begin{figure}
\captionsetup[subfloat]{farskip=0pt,captionskip=1pt}
{%
  \includegraphics[clip,width=\columnwidth]{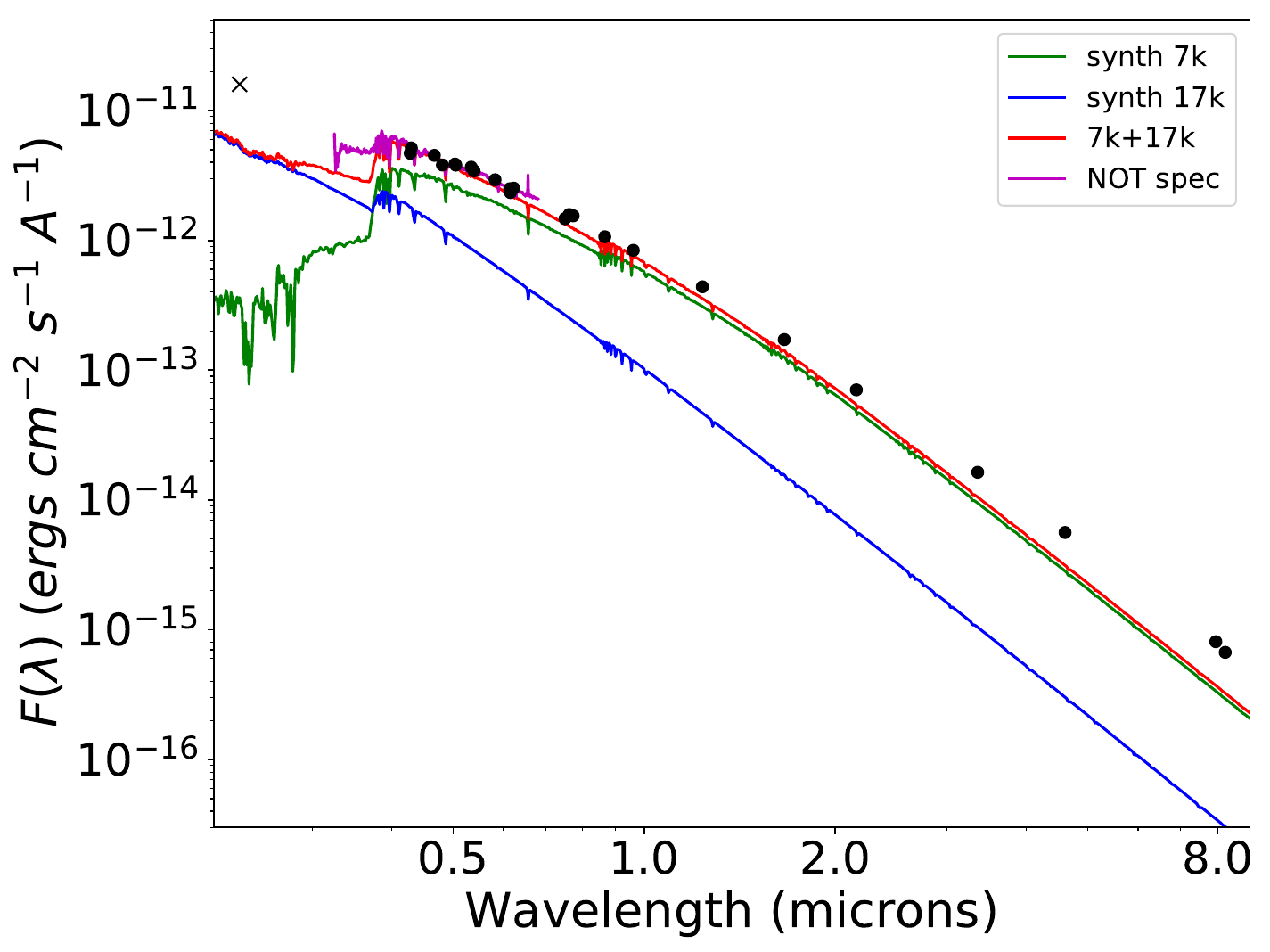}%
}
\caption{SED model for V1315 Cas using two templates with the derived \textsc{phoebe}2 parameters: one at 7\,000 K (green) and one at  17\,000 K (blue). The sum of these is shown in red and compared with the observational data (black dots) and the observed flux-calibrated spectrum (purple). The GALEX NUV point (black cross) was excluded from the analysis. The IR excess likely originates from a circumstellar and/or circumbinary disc.}
\label{fig:sed}
\end{figure}

We corrected all data in the SED analysis for an interstellar extinction of Av $\sim$ 2.7 (sec. \ref{sec:spec}). We fitted the data using Castelli-Kurucz stellar atmosphere models of two templates, using the parameters derived by the \textsc{phoebe}2 model (Table \ref{tab:phoebe}), and let the distance $d$ be a free parameter. We notice an IR excess, which we attribute to a circumstellar or circumbinary disc. We were not able to fit the \textit{GALEX} NUV point that was showing higher flux compared to the model or would require larger radius of the secondary which cannot be reproduced by the \textsc{PHOEBE2}’s model. We prefer at this stage to exclude it from our fit, but we show it in the figure. The FUV GALEX flux is not available. Apart from the NUV GALEX datapoint, all other values from photometry were considered. We derived a distance of 1.35 kpc, which is in a good agreement with the distance provided by Gaia EDR3 ($d=1.382$ kpc).  Fig. \ref{fig:sed} shows the energy distribution for each component and the combined spectrum along with the observed points and the NOT flux-calibrated low-resolution spectrum.

\section{Results and discussion}
The results from the \textsc{phoebe}2 modelling described in section \ref{phmod} indicate that the binary system consists of two components with a mass ratio $q=0.115$. The first component is an F-type evolved star while the other component is a B-type star. We list the parameters of both components in Table \ref{tab:phoebe}. Given that the less massive component is the most evolved, it is reasonable to assume that the system was or is still undergoing mass transfer and thus belongs to the class of interacting binaries. As we will show in the following section, the inflated radius of the primary star is the consequence of stellar evolution and the low mass  of the primary is due to the mass-transfer that removed the outer (hydrogen rich) layers of the primary.

\subsection{Mass ratio of the system}
\label{nat}

Interacting semi-detached binaries are often in synchronous rotation.
The derived radius of the primary is 18.51 R$_\odot$, while the measured rotational velocity is 25.8 $\pm$ 1.9 km/s. Assuming synchronous rotation, the estimated orbital period is 36.5 $\pm$ 2.6 days which is in good agreement with the observed period of 34.5 days.

Taken at face values, the primary is filling over 96 \% of its Roche lobe, but we cannot reject the hypothesis that it is completely filling it and that mass transfer is still ongoing. 

The mass ratio derived from the gaussian fitting of the helium I line gives a mass ratio of $0.35 \pm 0.11$. This is higher compared to the spectral disentangling method and the Phoebe model, however as discussed in sect. \ref{hellines} the helium I lines show large line profile variations.
The spectral disentangling method (sec \ref{specdis}) yields a mass ratio $q$ of $0.167 \pm 0.07$ from H$\beta$ line and $0.194 \pm 0.13$ from helium I lines. These results are compatible with the mass ratio derived by \textsc{Phoebe} of 0.115.

\subsection{Binary evolution}

V1315 Cas is likely an example of an Algol type system \citep{hil01} where the primary was originally more massive but after exhausting hydrogen in its core (and leaving the main-sequence) and filling its Roche-lobe due to stellar evolution it transferred a significant amount of mass through L1 point onto a companion which was originally less massive and less evolved. 

To explore the history and evolution of the system we have explored the model grids of BPASS \citep{eld17}. We were not able to find a solution for the observed properties. We have tried to primarily match the masses of both components and the temperature and radius of the primary as the temperature and radius of the secondary are constrained with higher uncertainties.

\begin{figure}
 \includegraphics[width=\columnwidth]{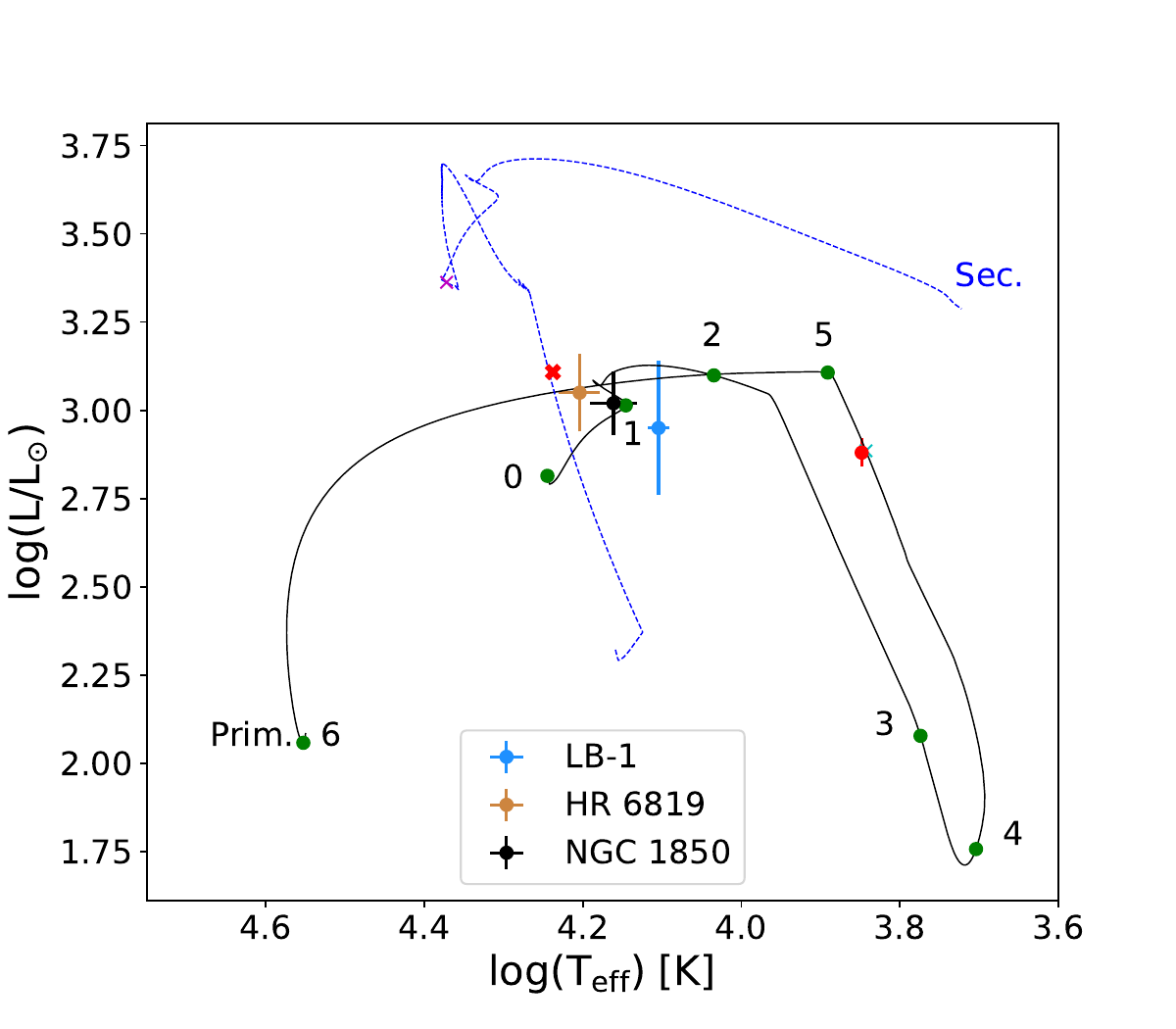}
 \caption{Evolutionary tracks for the best matching MESA model, with initial masses of 5.8 and 3.2 M$_\odot$: black is the primary and blue is the secondary. The green points mark important evolution points in the lifetime of the primary star, while the red circle corresponds to the position of the primary obtained from our analysis and overlaps a cyan cross which corresponds to the current position of primary from the evolutionary model. The red and purple crosses correspond to the position of the secondary obtained from our analysis and evolutionary models, respectively.}
 \label{fig:hrd}
\end{figure}

We have thus decided to use the Modules for Experiments in Stellar Astrophysics \citep[MESA,][]{pax11,pax15} to construct a very simple evolutionary model. We have made use of the mass transfer prescription of \citet{kolb90}. As a first attempt to use scenarios based on fully conservative mass transfer didn't allow us to find any solution that matches the observed properties, we generated different starting models with varying initial masses of both components, initial period and $\alpha, \beta, \gamma, \delta $ coefficients of non-conservative mass transfer as described in section 16.4.1 of \citet{tau06}. Our best match model starts with initial masses of 5.8 and 3.2~M$_\odot$,  a 6.95 d orbital period, and values of $\alpha, \beta=0$, $\delta=0.2$ and $\gamma=1.2$, meaning that 20\% of the material lost from the primary (that is, $\approx1$ M$_\odot$) is not accreted by the secondary and that there is a circumbinary coplanar toroid with radius $\gamma^2 a = 1.44 a$.
We present the evolution of various stellar parameters in Figures \ref{fig:hrd}, \ref{fig:ETs1}, \ref{fig:ETs2}, and \ref{fig:ETs3}. The stellar parameters from the evolutionary model are in a good agreement to the ones from our analysis. The most discrepant parameter between the evolutionary models and our analysis is the current temperature of the secondary. However, this parameter is likely affected by accretion processes.

We have highlighted and labeled important evolution points of the primary in Fig. \ref{fig:hrd}. The stellar evolution starts with initial masses of 5.8 M$_\odot$ of the primary (point 0) and 3.2 M$_\odot$ of the secondary. 

After the primary has exhausted hydrogen in its core (1), the luminosity increases due to hydrogen shell burning. Shortly after, mass transfer from the primary to the secondary begins (2), which slightly decreases the orbital period as the mass losing star is the more massive one. Point (2) is also where the binary track of the primary star deviates from a single star evolution scenario. After the inversion of the mass ratio (3), the mass transfer continues, but now the orbit is significantly widening as the mass losing star is already the less massive one. The maximum rate of mass transfer (4) is $\sim$\SI{7.71e-5} M$_\odot$/yr. After reaching the lowest luminosity, the primary starts to expand again and continues to the region where it is observed today, with an age of $\sim$\SI{7.7e7} years. The mass transfer ends (5) and the system evolves further as a detached one. Our model ends (6) when the primary has depleted its helium and reached an effective temperature of $\approx 35\,700$ K.
 The obtained chemical abundances (section \ref{abund}) are in qualitative agreement with the abundances obtained from our MESA model as can be seen in Fig. \ref{fig:ETs2}. Once the mass-transfer started and removed the outer layers of the primary, the inner layers with CNO-processed abundances were exposed.
We acknowledge that the found model is rather a representative example of a possible evolutionary channel rather than being an unique solution.

Currently, the system is in a similar evolutionary phase as HD 15124 \citep{elbad22} and V393 Sco \citep{menn12a}: systems that have already inverted their mass ratio but have had not enough time to detach. These systems may evolve into a classical Be star with a decretion disc and a stripped sdOB companion. We are unable to accurately measure the rotational velocity of the secondary and thereby infer whether the secondary might eventually form a decretion disc. 

In the literature we have found several examples of systems in similar evolutionary state \citep[e.g.][]{barr13,har15,ros18,elb_apo}. We found that V495 Cen \citep{ros18} resembles the studied system the most except that it is an eclipsing one. Furthermore, data from Gaia DR3 yielded 14 similar systems with high mass function \citep{EB22gaia}.

The orbital period is predicted to reach 38.9 days after the detachment. The radius of the primary will decrease while its temperature will increase and will radiate most of its energy in the UV region. The system will resemble HR 6819 \citep{bod20}, LB-1 \citep{she20}, NGC 1850 \citep{sara23} and other stripped star binaries systems that were initially reported to contain a dormant black hole.

\section{Summary}
This paper presents a study of a newly discovered interacting non-eclipsing binary V1315 Cas. Using photometry, high-resolution spectroscopy and broad-band SED we show that system is not a dormant black hole, but that it is a wide binary ($P_{\rm{orb}}=34.53$ d) on a nearly circular orbit that consists of an F-type evolved star that underwent mass transfer and a B-type companion star.

We derived the properties of the primary, mass $\rm{M_p}= 0.84$ M$_\odot$, radius $\rm{R_p}=18.51$ R$_\odot$, and effective temperature $\rm{T_p}=7\,050$ K, and of the secondary $\rm{M_s}= 7.3$ M$_\odot$, $\rm{R_s}=4.02$ R$_\odot$,  $\rm{T_s}=17\,290$\,K.

We detect significant IR excess alongside of strong DIBs and also UV-excess. Balmer lines show a complex character composed of several absorption and emission components.

We investigated the evolutionary history of the system with MESA. The best model that we found assumes that the system started with a primary initial mass of 5.8 M$_\odot$, a secondary mass of 3.2 M$_\odot$, and an initial orbital period of $P_{\rm{orb}}=6.95$~d. After $\sim$\SI{7.69e7} years a rapid non-conservative mass transfer started which led to an inversion of the mass ratio and an increase of the orbital period that we observe. Furthermore, the stationary H$\alpha$ emission component likely originates from the circumbinary toroid as found by our best MESA model.

We detect strong nitrogen enhancement and carbon depletion. The derived abundances can be explained by CNO processed material reaching the surface of the primary due to mass transfer exposing the inner layers. After the mass transfer stage ends the system will resemble stripped star binaries such as HR 6819 and LB-1. Such systems consist of a stripped star that will become a sdOB star.

Future high S/N optical spectra will shed more light into the peculiar behavior of helium lines displaying asymmetries and periodicity. Future ultraviolet data could provide necessary information to fully characterize the presence of the accretion disc.

Future missions like Polstar \citep{jon21, pet21} and UVEX \citep{kul21}, working in synergy with large scale surveys such as WEAVE, LAMOST, 4MOST, and others, will allow discovering several thousands of stripped stars and for a few selected ones, we should be able to characterize their binary evolution.

\section*{Acknowledgements}
The authors would like to thank the anonymous referee for their insightful report.
The authors thankfully acknowledge the technical expertise and assistance provided by the Spanish Supercomputing Network (Red Espa\~nola de Supercomputaci\'on), as well as the computer resources used: the LaPalma Supercomputer, located at the Instituto de Astrof\'isica de Canarias.  This research was supported by the  Erasmus+  programme  of the European Union under grant numbers 2017-1-CZ01-KA203-035562 and 2020-1-CZ01-KA203-078200.

This paper includes data collected by the TESS mission. Funding for the TESS mission is provided by the NASA Explorer Program.  This research was based on observations obtained with the HERMES spectrograph, which is supported by the Fund for Scientific Research of Flanders (FWO), Belgium, the Research Council of KU Leuven, Belgium, the Fonds National de la Recherche Scientifique (FNRS), Belgium, the Royal Observatory of Belgium, the Observatoire de Geneve, Switzerland, and the Thuringer Landessternwarte Tautenburg, Germany. The Mercator telescope is operated thanks to grant number G.0C31.13 of the FWO under the ‘Big Science’ initiative of the Flemish government.

Based on observations made with the Nordic Optical Telescope, owned in collaboration by the University of Turku and Aarhus University, and operated jointly by Aarhus University, the University of Turku and the University of Oslo, representing Denmark, Finland and Norway, the University of Iceland and Stockholm University at the Observatorio del Roque de los Muchachos, La Palma, Spain, of the Instituto de Astrof\'sica de Canarias. The data presented here were obtained in part with ALFOSC, which is provided by the Instituto de Astrofisica de Andalucia (IAA) under a joint agreement with the University of Copenhagen and NOTSA.

DJ acknowledges support from the Erasmus+ programme of the European Union under grant number 2020-1-CZ01-KA203-078200. HVW acknowledges support from the Research Council of the KU Leuven under grant number C14/17/082.
MS acknowledges the support by Inter-transfer grant no LTT-20015.
PGB acknowledges support by the Spanish Ministry of Science and Innovation with the \textit{Ram{\'o}n\,y\,Cajal} fellowship number RYC-2021-033137-I and the number MRR4032204. TS acknowledges support from the European Union's Horizon 2020 Marie Skłodowska-Curie grant No. 101024605.

This work made use of \texttt{tpfplotter}, which also made use of the python packages \texttt{astropy}, \texttt{lightkurve}, \texttt{matplotlib} and \texttt{numpy}. 

\section*{DATA AVAILABILITY}

The data underlying this article will be shared on reasonable request to the corresponding author.



\bibliographystyle{mnras}
\bibliography{NEW2} 




\appendix
\section{Appendix}
\renewcommand\thefigure{\thesection.\arabic{figure}} 

\setcounter{figure}{0}   

\begin{table*}
\begin{center}
\centering
\caption{Radial velocity measurements of V1315 Cas. The columns indicate the MJD, the exposure time, the telescope used (TAU=TLS; MER=Mercator), the radial velocity of the primary star, and the radial velocity of the secondary from fitting the helium I line (in a few spectra the fit was not satisfactory and this last column was left blank).}

\begin{tabular}{ r c c r r}
\hline
MJD (days) & Exp. time (s) & Telescope& $\rm{RV_{Prim}}$ (km/s) &$\rm{RV_{Sec}}$ (km/s)  \\
\hline
58389.14910  &1200 & TAU &  135.76  $\pm$ 0.19&  -8.6 $\pm$ 7.4 \\
58836.86788  &1000 & MER &  120.36  $\pm$ 0.13&  -49.7 $\pm$ 7.6 \\
58851.83548  &1000 & MER &  4.88   $\pm$ 0.14&  -29.9 $\pm$ 5.9 \\
58853.84162  &1000 & MER &  -27.87  $\pm$ 0.15&  -29.6 $\pm$ 7.1 \\
58873.87366  &1000 & MER &  144.85  $\pm$ 0.14&  -24.4 $\pm$ 5.8 \\
58876.85141  &1000 & MER &  146.47  $\pm$ 0.14&  23.00 $\pm$ 6.4 \\
58887.83714  &900 & MER  &  -19.88  $\pm$ 0.14&  -28.2     $\pm$ 7.1 \\
58896.83510  &1100 & MER &  -39.87  $\pm$ 0.13&  -62.5 $\pm$ 7.9 \\
59045.21759  &1200 & MER &  134.73  $\pm$ 0.12&  -31.6 $\pm$ 7.0 \\
59065.09281  &1200 & MER &  -64.45  $\pm$ 0.11&   \\
59065.24312  &600 & MER  &  -64.90  $\pm$ 0.16&   \\
59066.08154  &900 & MER  &  -64.94  $\pm$ 0.14&  -79.3 $\pm$ 6.9 \\
59066.22751  &1200 & MER &  -64.97  $\pm$ 0.11&   \\
59067.23124  &1200 & MER &  -61.61  $\pm$ 0.13&   \\
59068.23311  &1200 & MER &  -54.53  $\pm$ 0.12&   \\
59069.24078  &1200 & MER &  -43.85  $\pm$ 0.13&  -78.3 $\pm$ 6.0 \\
59070.18420  &1200& MER  &  -32.04  $\pm$ 0.13&  -73.5 $\pm$ 6.4 \\
59071.22796  &900 & MER  &  -15.77  $\pm$ 0.15&  -63.41 $\pm$ 6.6 \\
59072.21963  &1200 & MER &  2.27    $\pm$ 0.13&   -47.3   $\pm$ 7.1 \\
\hline 
\end{tabular}
\label{tab:rv}
\end{center}
\end{table*}

\begin{figure}
\captionsetup[subfloat]{farskip=0pt,captionskip=1pt}
{%
  \includegraphics[clip,width=\columnwidth]{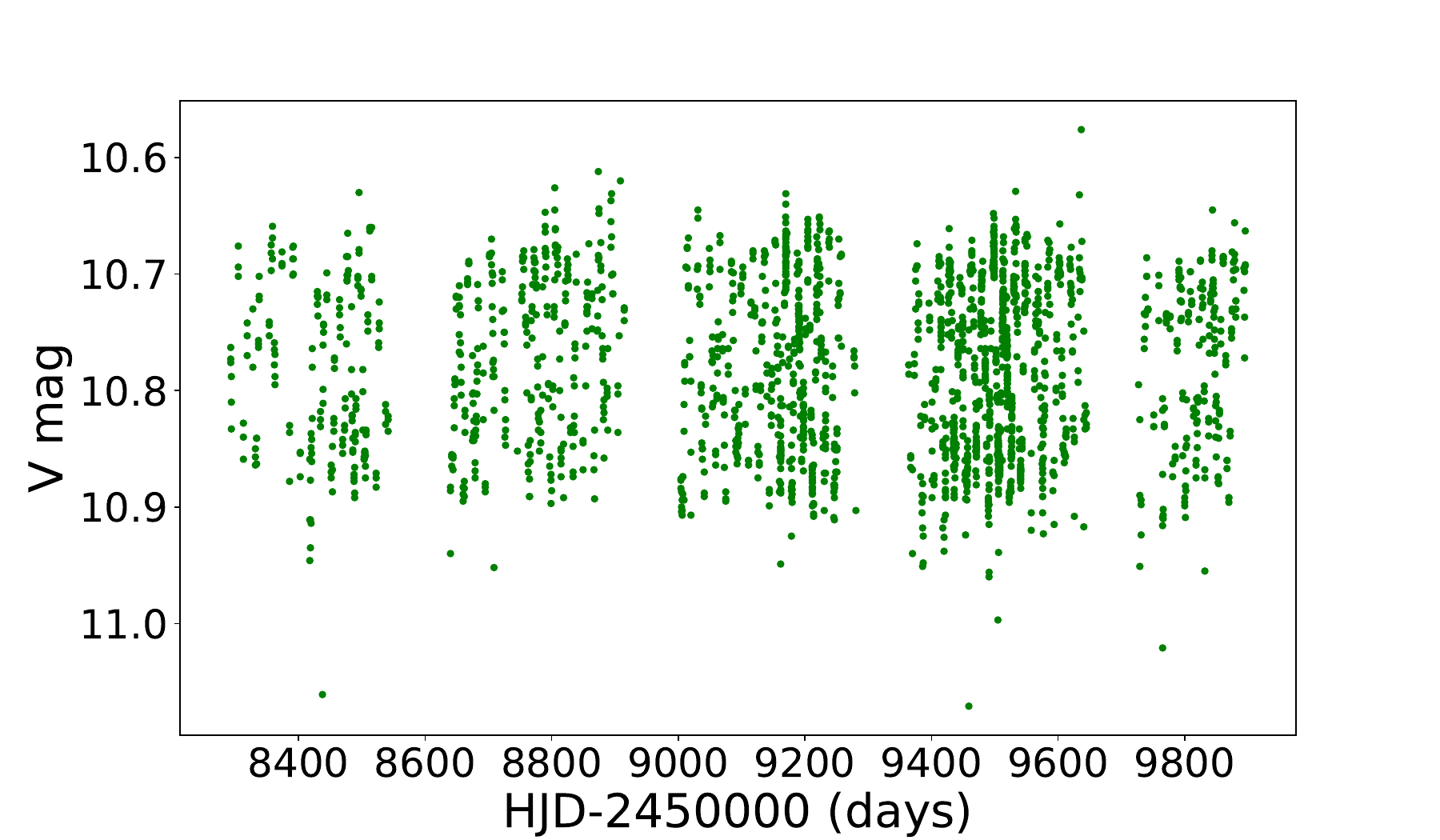}%
}

{%
  \includegraphics[clip,width=\columnwidth]{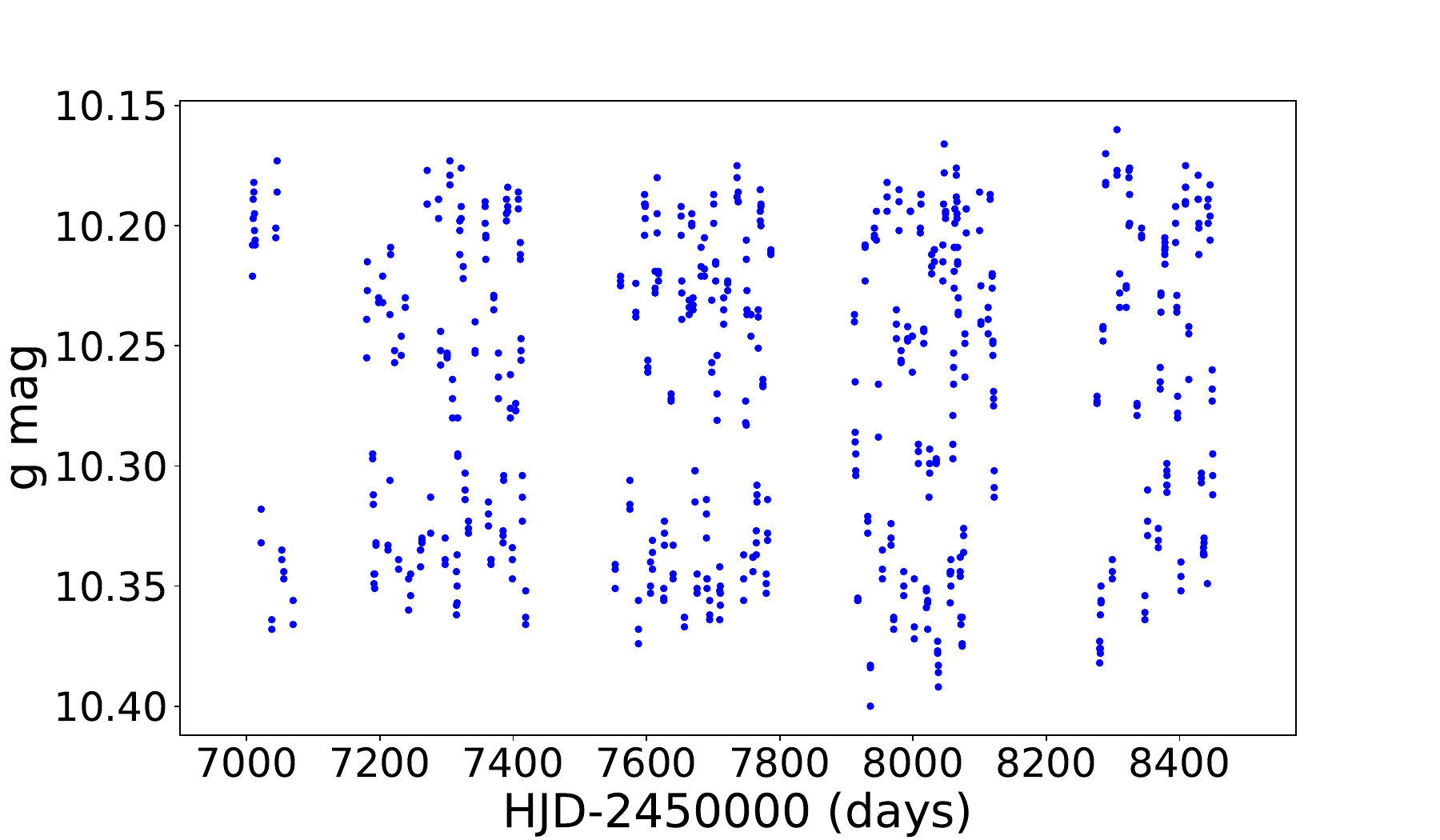}%
}

\caption{The unfolded light curves from ASAS-SN. Data in \textit{V} and \textit{g} band are in top and bottom subfigure respectively.}
\label{fig:lcun}
\end{figure}

\begin{figure}
\captionsetup[subfloat]{farskip=0pt,captionskip=1pt}
{%
  \includegraphics[clip,width=\columnwidth]{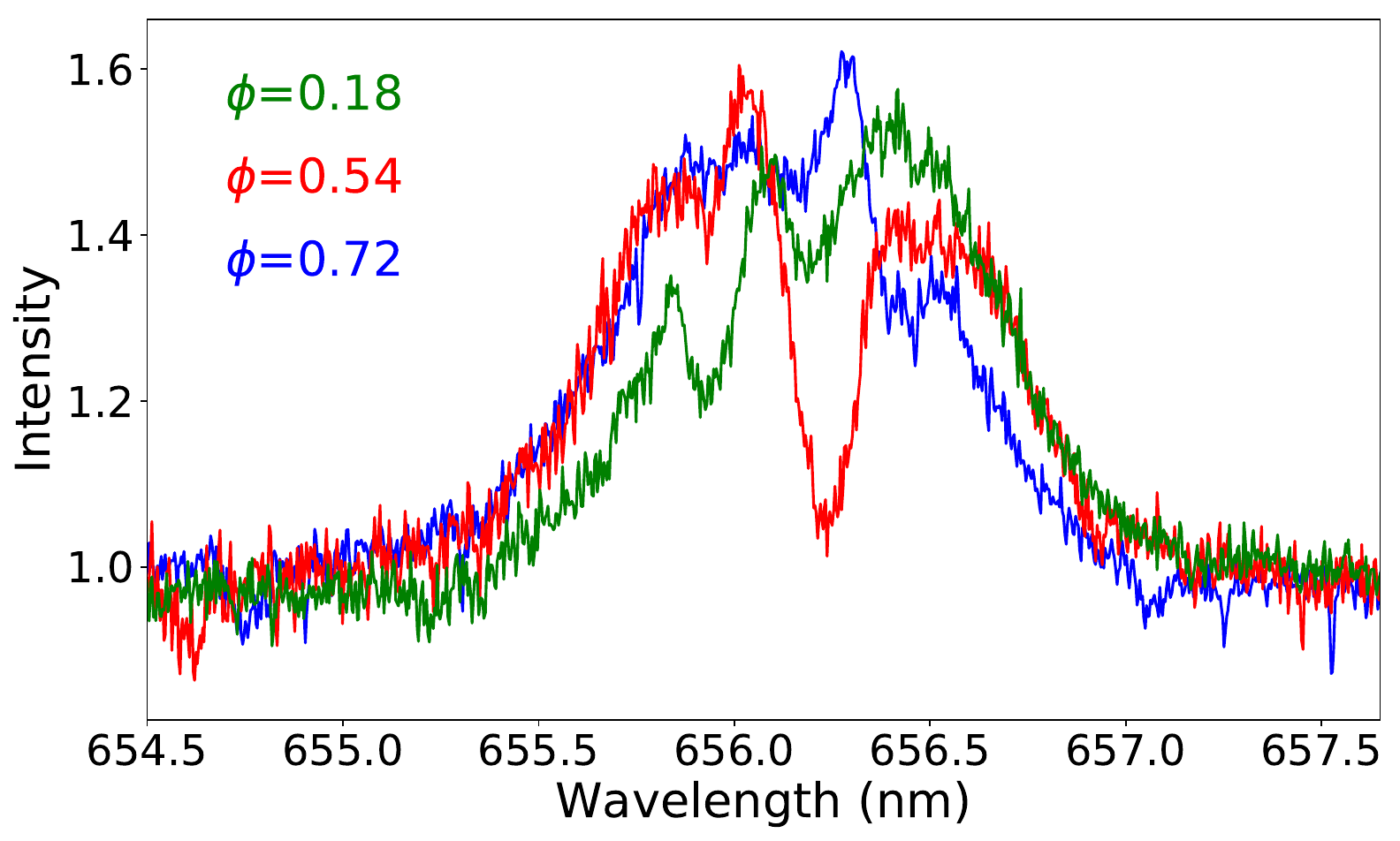}%
}

{%
  \includegraphics[clip,width=\columnwidth]{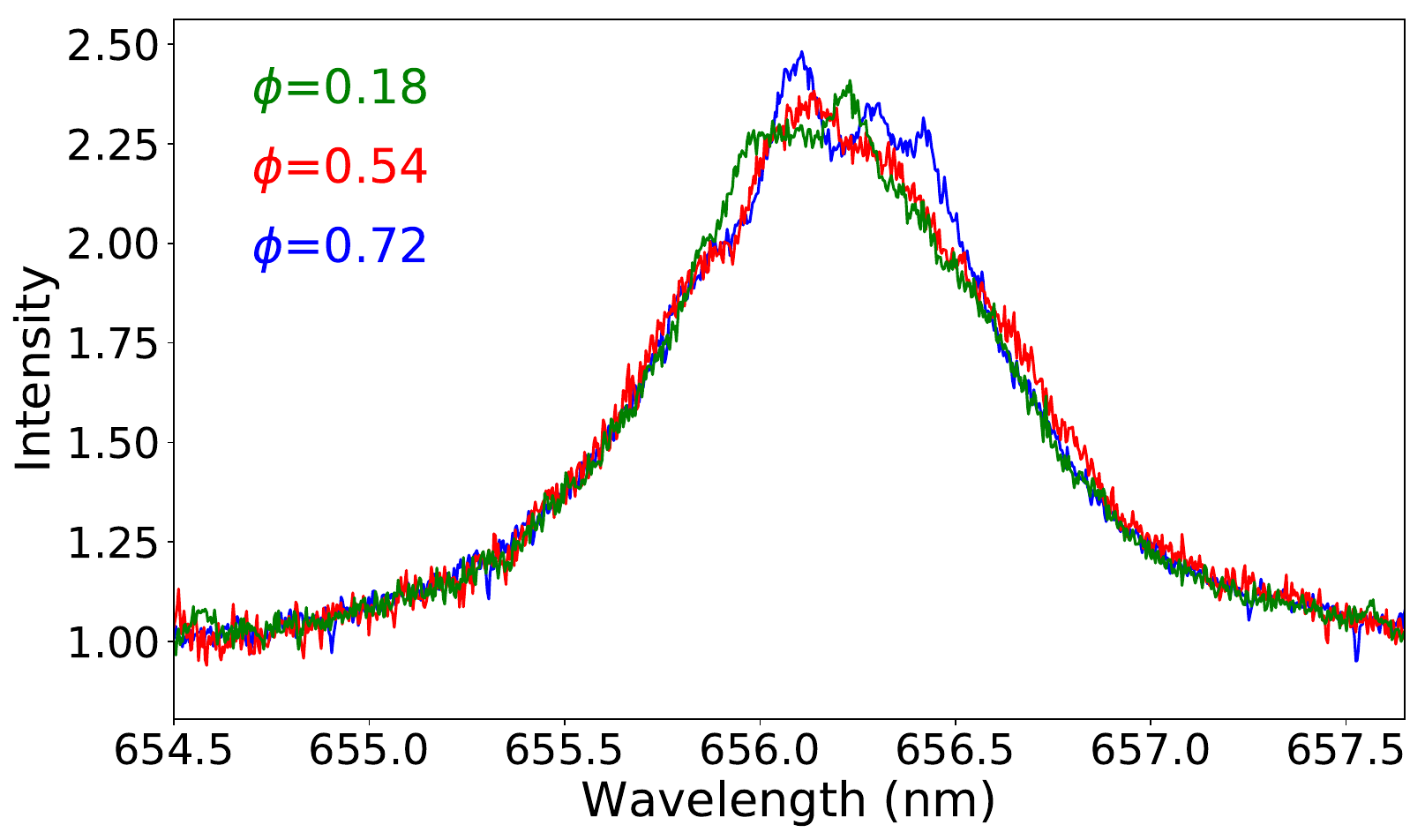}%
}

\caption{Complex structure of the H$\alpha$ line. Three spectra taken at different orbital phases (differentiated by color). \textit{Top:} Observed spectra. \textit{Bottom:} Spectra after subtracting two absorption components. We see that the emission wings show no significant motion during the orbital phase. Large error bars on the center position do not claim a static emission component rather limit the motion of the emission component.}
\label{fig:hal}
\end{figure}

 \begin{figure*}
 \includegraphics[width=\textwidth]{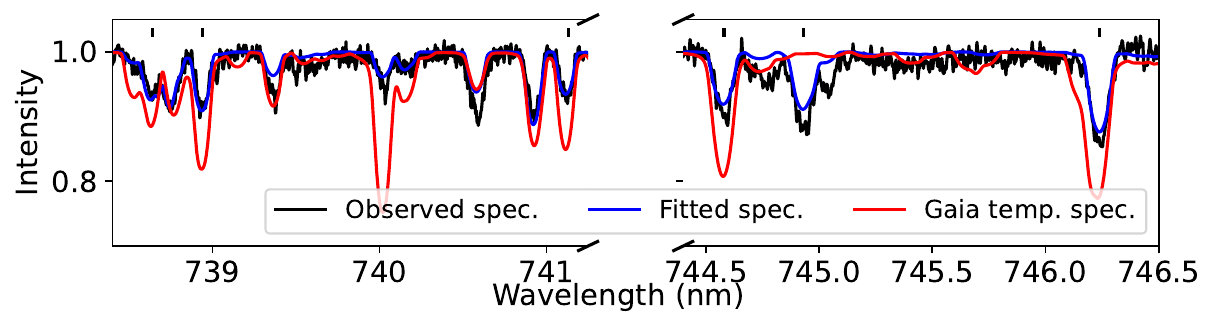}
 \caption{Region of the observed Mercator spectrum is shown in black. We compare the spectrum we derived in our analysis with $T_{\rm{eff}}=7000$ K (blue) to the spectrum with the temperature from Gaia DR2 with $T_{\rm{eff}}=4376$ K (red) clearly showing the later one is not reliable. With short black vertical lines we mark the positions of the iron lines used for the analysis. }
 \label{speccomp}
\end{figure*}

 \begin{figure*}
 \includegraphics[width=\textwidth]{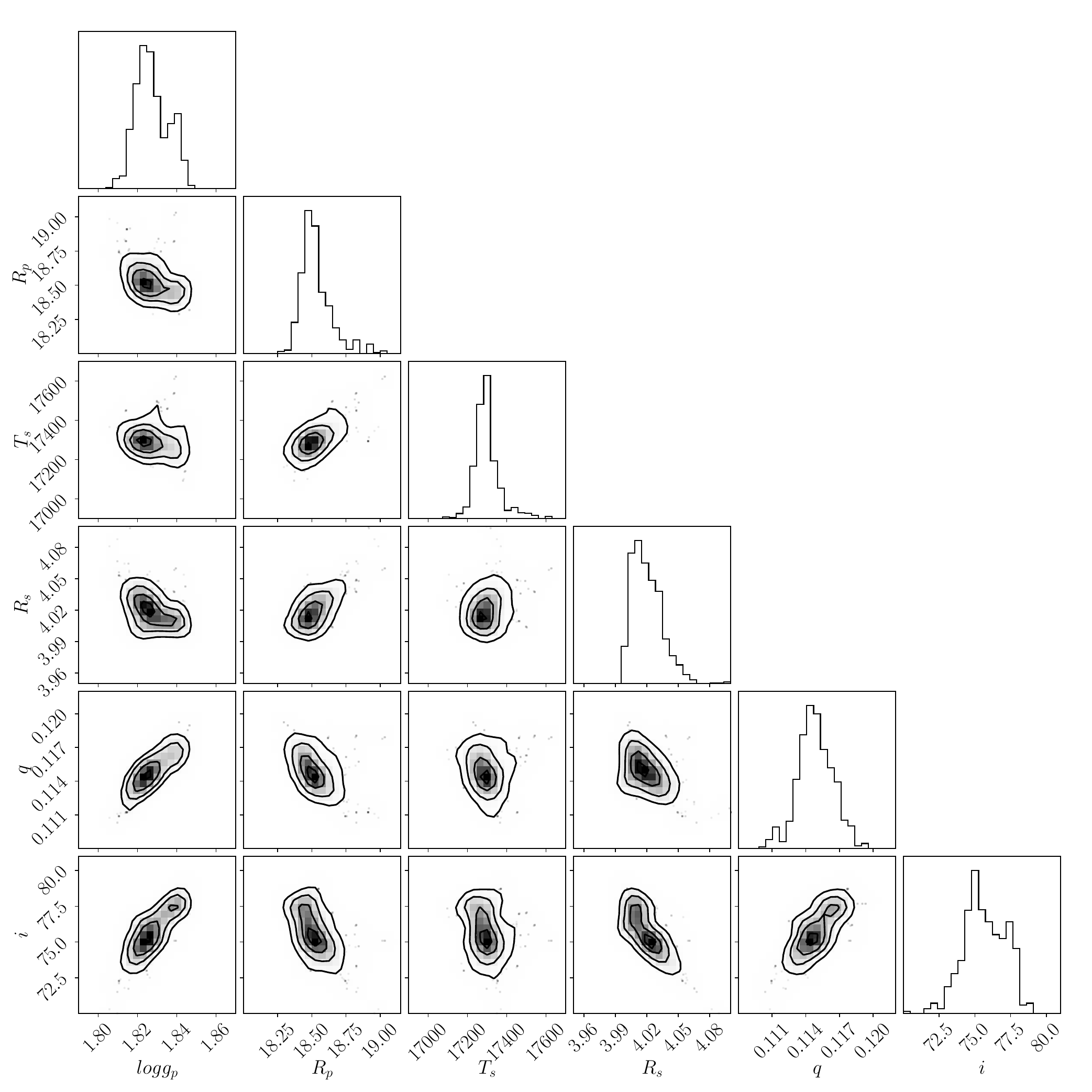}
 \caption{Corner plot of the MCMC posteriors from the \textsc{phoebe}2 fitting. }
 \label{fig:mcmc}
\end{figure*}

\begin{figure}
\captionsetup[subfloat]{farskip=0pt,captionskip=1pt}
{%
  \includegraphics[clip,width=\columnwidth]{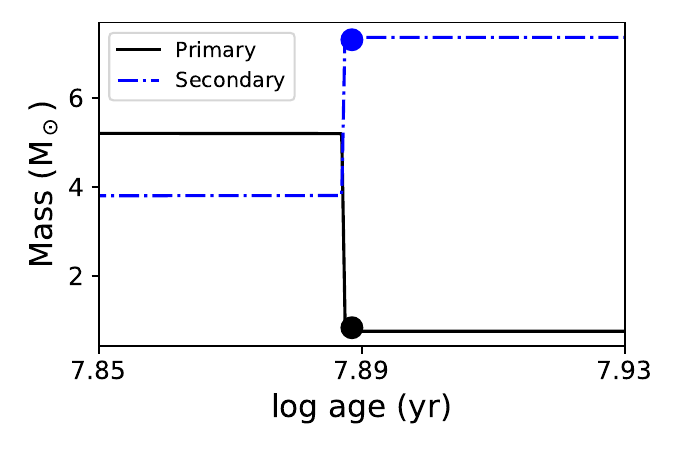}%
}

{%
  \includegraphics[clip,width=\columnwidth]{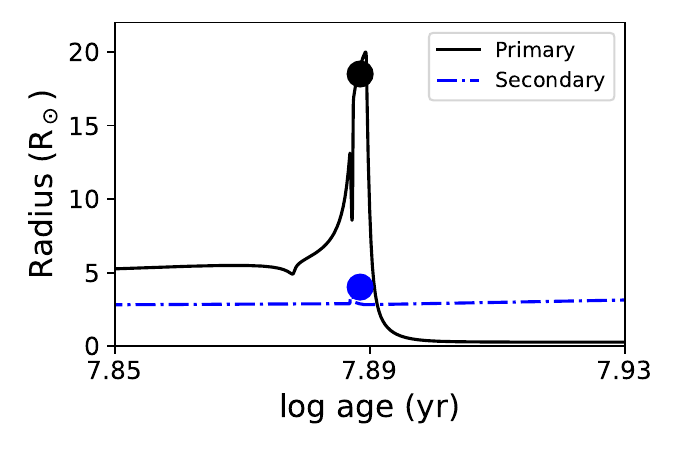}%
}
\caption{Physical parameters over the evolution of the binary system.  The values we derived in this paper are marked with circles. \textit(Top) Mass evolution of both components; \textit(Bottom) Radius evolution of both components.}
\label{fig:ETs1}
\end{figure} 

\begin{figure}
\captionsetup[subfloat]{farskip=0pt,captionskip=1pt}
{%
  \includegraphics[clip,width=\columnwidth]{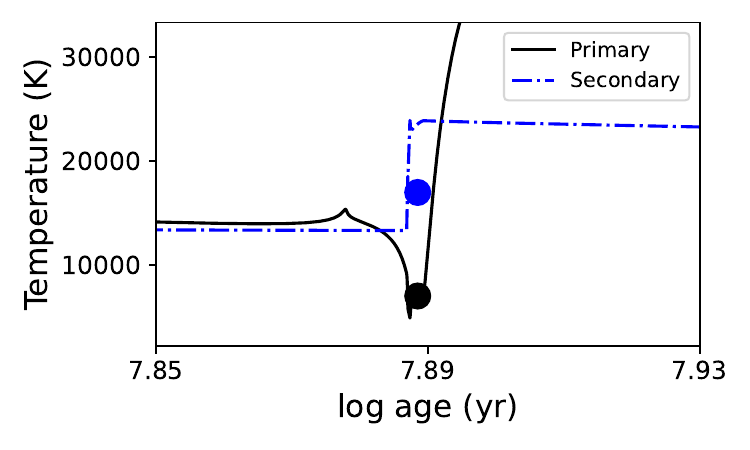}%
}

{%
  \includegraphics[clip,width=\columnwidth]{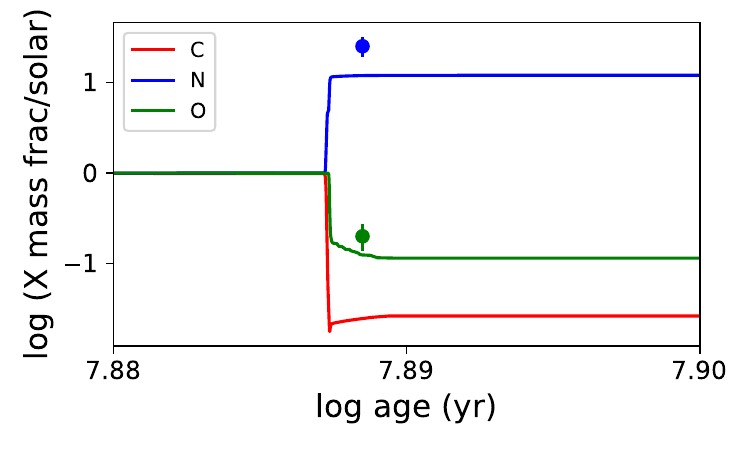}%
}

\caption{Evolution of the temperature of both components (top) and of the CNO abundances of the primary, as obtained with our best fit MESA model. The measured N and O abundances are shown. Carbon is depleted.}
\label{fig:ETs2}
\end{figure} 

\begin{figure}
\captionsetup[subfloat]{farskip=0pt,captionskip=1pt}
{%
  \includegraphics[clip,width=\columnwidth]{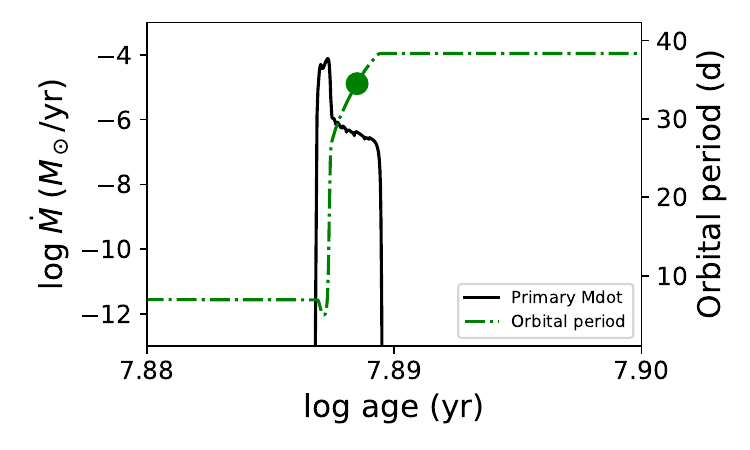}%
}

\caption{Evolution of the orbital period (right axis) and mass loss (left axis) of the primary. Due to angular momentum transfer, the orbit has expanded from an initial period of 6.95 days to the current 34.5 days. }
\label{fig:ETs3}
\end{figure} 
\bsp	
\label{lastpage}
\end{document}